\documentclass[aos, preprint]{imsart}

\usepackage{amsthm, amsmath, amssymb}
\usepackage{natbib}
\usepackage[boxed]{algorithm2e}
\usepackage{graphicx}
\usepackage{enumerate}

\graphicspath{{./spCIsimulations/graphics/}{./spLAanalysis/graphics/}}
\DeclareGraphicsExtensions{.pdf}

\startlocaldefs
    \newcommand{\indicator}[1]{\ensuremath{\mathbf{1}_{#1}}}
    \newcommand{\dif}[1][\:]{\ensuremath{#1 \mathrm{d}}}
    
    \newcommand{\E}[2][]{\ensuremath{\mathbb{E} #1[\, #2 \, #1]}}
    \newcommand{\condE}[3][]{\ensuremath{\mathbb{E} #1[\, #2 \, #1|\, #3 \, #1]}}
    \newcommand{\subE}[3][]{\ensuremath{\mathbb{E}_{#2} #1[\, #3 \, #1]}}
    \newcommand{\Var}[2][]{\ensuremath{\mathrm{Var} #1( #2 #1)}}
    
    \newcommand{\subVar}[3][]{\ensuremath{\mathrm{Var}_{#2} #1( #3 #1)}}
    \newcommand{\Cov}[3][]{\ensuremath{\mathrm{Cov} #1( #2, #3 #1)}}
    \newcommand{\subCov}[4][]{\ensuremath{\mathrm{Cov}_{#2} #1( #3, #4 #1)}}
    
    \newcommand{\deriv}[1]{\ensuremath{\frac{\partial}{\partial #1}}}
    \newcommand{\secderiv}[1]{\ensuremath{\frac{\partial^2}{\partial #1^2}}}
    \newcommand{\dist}{\ensuremath{\stackrel{\text{D}}{\longrightarrow}}}
    \newcommand{\prob}{\ensuremath{\stackrel{\text{P}}{\longrightarrow}}}
    \newcommand{\trans}{\ensuremath{\mathsf{T}}}

    \newcommand{\Inf}{\ensuremath{\mathcal{I}}}

\endlocaldefs

\begin{document}

\begin{frontmatter}
    \title{Semiparametric Relative-risk Regression for Infectious Disease Data}
    \runtitle{Regression for Infectious Disease Data}

    \author{\fnms{Eben} \snm{Kenah} \thanksref{t1} \ead[label=e1]{ekenah@ufl.edu}}
    \thankstext{t1}{The author is grateful for the comments of M. Elizabeth Halloran and Ira M. Longini, Jr. The Los Angeles County Department of Public Health generously allowed the use of their data in Section 4. This research was supported by National Institute of Allergy and Infectious Diseases (NIAID) grant K99/R00 AI095302. The content is solely the responsibility of the author and does not necessarily represent the official views of NIAID or the National Institutes of Health.}
    \address{22 Buckman Drive, Dauer Hall\\ PO Box 117450\\ Gainesville, FL 32611-7450\\ \printead{e1}}
    \affiliation{Biostatistics Department and Emerging Pathogens Institute,\\ University of Florida}
    \runauthor{E. Kenah}

    \begin{abstract}
    This paper introduces semiparametric relative-risk regression models for infectious disease data based on contact intervals, where the contact interval from person $i$ to person $j$ is the time between the onset of infectiousness in $i$ and infectious contact from $i$ to $j$.  The hazard of infectious contact from $i$ to $j$ is $\lambda_0(\tau)r(\beta_0^\trans X_{ij})$, where $\lambda_0(\tau)$ is an unspecified baseline hazard function, $r$ is a relative risk function, $\beta_0$ is an unknown covariate vector, and $X_{ij}$ is a covariate vector. When who-infects-whom is observed, the Cox partial likelihood is a profile likelihood for $\beta$ maximized over all possible $\lambda_0(\tau)$.  When who-infects-whom is not observed, we use an EM algorithm to maximize the profile likelihood for $\beta$ integrated over all possible combinations of who-infected-whom. This extends the most important class of regression models in survival analysis to infectious disease epidemiology.
    \end{abstract}


\end{frontmatter}

\section{Introduction}
\label{sec:intro}
Infectious diseases remain an important threat to human health and commerce, and understanding the effects of covariates on infection transmission is crucial to the design of public health interventions. The statistical analysis of infectious disease data is complicated by the fact that the outcomes (infections) are inherently dependent~\citep{Becker, AnderssonBritton}. This problem is especially pronounced for diseases transmitted directly from person to person, such as influenza and SARS. Epidemiologists have dealt with this problem in three ways. Most commonly, they model susceptibility to disease using standard statistical methods, such as logistic or Cox regression, that ignore this dependence. A second approach is to use discrete-time chain binomial models~\citep{RampeyLongini} to estimate the probability of escaping infectious contact from infected members of close-contact groups such as households, classrooms, or hospital wards. The third and most recent approach is to model the spread of disease as a branching process where infectees are the offspring of their infectors~\citep{WallingaTeunis, WhitePagano}. The time intervals between the infection of an infector and the infection of his or her infectees are called generation intervals. Generation intervals in the branching process are assumed to be independent and identically distributed (iid) samples from a known or estimated distribution.

To understand transmission, it is crucial to separate the effects of covariates on the risk of transmission (i.e., infectiousness and susceptibility) from their association with exposure to infected people~\citep{Rhodes}. Chronic-disease models cannot do this, inherently conflating susceptibility and exposure. At the other extreme, generation and serial interval methods model the transmission of disease as a process that creates a population, not a process of spread through a preexisting population. Since infected people are the offspring of an infectious parent, these models ignore competing risks of infection from multiple infectors~\citep{Svensson, Kenah3}. Since the generation intervals are iid, these models force the implicit assumption of constant latent and infectious periods~\citep{Kenah5} and cannot be extended easily to model covariate effects. Discrete-time chain binomial models are a statistically sound response to the problem of dependence, and they separate the effects of covariates on the risk of transmission from their association with exposure to infectious persons. However, their use is limited in two ways: First, they are not implemented in standard statistical software---a problem solved partially by the publicly-available package TranStat (\texttt{www.epimodels.org/midas/transtat.do}). Second, they force the use of discrete time. Since infectious disease data are usually recorded by the day or week, this is not unnatural. However, continuous-time models corrected for ties may offer a more natural and more flexible modeling framework.

\citet{Kenah4} showed that parametric methods from survival analysis could be extended to analyze infectious disease data by modeling the contact interval. The contact interval $\tau_{ij}$ in the ordered pair $ij$ is the time between the onset of infectiousness in $i$ and the first infection contact from $i$ to $j$, where infectious contact is defined as a contact sufficient to infect a susceptible individual. It is right-censored if the infectious period of $i$ ends before $i$ makes infectious contact with $j$ or if $j$ is infected by someone other than $i$. The distribution of $\tau_{ij}$ provides a concise summary of the evolution of infectiousness over time in person $i$ because its hazard function equals the hazard of infectious contact with $j$. These methods solve the problem of dependence by treating pairs of individuals, not the individuals themselves, as the units of analysis. \citet{Kenah5} showed that the contact interval distribution could be estimated nonparametrically by extending the Nelson-Aalen estimator from standard survival analysis. These methods assume a homogeneous population which the contact interval distribution is the same for all ordered pairs $ij$ where transmission from $i$ to $j$ is possible. As such, they are unable to address many important questions in infectious disease epidemiology.

The goal of this paper is to extend the nonparametric estimators in~\citet{Kenah5} to obtain a relative-risk regression model similar to that of~\citet{Cox1972} that will allow the semiparametric estimation of the effects of covariates on the hazard of infectious contact. For the ordered pair $ij$, the covariate vector can include infectiousness covariates for $i$, susceptibility covariates for $j$, and pairwise covariates. 

The rest of Section~\ref{sec:intro} reviews nonparametric estimation of the contact interval distribution, and Section~\ref{sec:methods} extends this to relative risk regression models. Our derivations are based on counting processes and martingales. Good introductions to these ideas are given in~\citet{KalbfleischPrentice} and~\citet{Aalen}; \citet{FlemingHarrington} and~\citet{ABGK} have more detailed discussions. Section~\ref{sec:sims} explores the performance of the regression models in simulations, and Section~\ref{sec:data} uses them to analyze data from Los Angeles County during the 2009 influenza A(H1N1) pandemic. Section~\ref{sec:discussion} discusses the promise and peril of semiparametric relative risk regression in infectious disease epidemiology.

\subsection{Stochastic S(E)IR epidemic model}
\label{sec:seir}
Consider a closed population of $n$ individuals assigned indices $1\ldots n$.  Each individual is in one of four states: susceptible (S), exposed (E), infectious (I), or removed (R).  Person $i$ moves from S to E at his or her \textit{infection time} $t_i$, with $t_i = \infty$ if $i$ is never infected.  After infection $i$ has a \textit{latent period} of length $\varepsilon_i$, during which he or she is infected but not infectious.  At time $t_i +\varepsilon_i$, $i$ moves from E to I, beginning an \textit{infectious period} of length $\iota_i$.  At time $t_i+ \varepsilon_i + \iota_i$, $i$ moves from I to R.  Once in R, $i$ can no longer infect others or be infected.  The latent period is a nonnegative random variable, the infectious is a strictly positive random variable, and both have finite mean and variance.

An epidemic begins with one or more persons infected from outside the population, which we call \textit{imported infections}.  For simplicity, we assume that epidemics begin with one or more imported infections at time $0$ and there are no other imported infections.

After becoming infectious at time $t_i + \varepsilon_i$, person $i$ makes infectious contact with $j\neq i$ at time $t_{ij} = t_i + \varepsilon_i + \tau^*_{ij}$, where the \textit{infectious contact interval} $\tau^*_{ij}$ is a strictly positive random variable with $\tau^*_{ij} = \infty$ if infectious contact never occurs.  Since infectious contact must occur while $i$ is infectious or never, $\tau^*_{ij} \in (0, \iota_j]$ or $\tau^*_{ij} = \infty$.  We define infectious contact to be sufficient to cause infection in a susceptible person, so $t_j\leq t_{ij}$.

For each ordered pair $ij$, let $C_{ij} = 1$ if infectious contact from $i$ to $j$ is possible and $C_{ij} = 0$ otherwise.  We assume that the infectious contact interval $\tau^*_{ij}$ is generated in the following way: A \textit{contact interval} $\tau_{ij}$ is drawn from a distribution with hazard function $\lambda_{ij}(\tau)$.  If $\tau_{ij}\leq\iota_i$ and $C_{ij} = 1$, then $\tau^*_{ij} = \tau_{ij}$.  Otherwise, $\tau^*_{ij} = \infty$.  In this paper, we assume the contact intervals in all ordered pairs $ij$ are independent and have finite mean and variance.  

Following \citet{WallingaTeunis}, let $v_j$ denote the index of the person who infected person $j$, with $v_j = 0$ for imported infections and $v_j = \infty$ for persons not infected at or before time $T$.  The \textit{transmission network} is the directed network with an edge from $v_j$ to $j$ for each $j$ such that $t_j\leq T$.  It can be represented by a vector $\mathbf{v} = (v_1,\ldots, v_n)$.  Let $\mathcal{V}_j = \{i: C_{ij}I_i(t_j) = 1\}$ denote the set of possible infectors of person $j$, which we call the \textit{infectious set} of person $j$.  Let $\mathcal{V}$ denote the set of all possible $\mathbf{v}$ consistent with the observed data. A $\mathbf{v}\in\mathcal{V}$ can be generated by choosing a $v_j\in\mathcal{V}_j$ for each non-imported infection $j$.

Our population has size $n$, and we observe the times of all $\text{S}\rightarrow\text{E}$ (infection), $\text{E}\rightarrow\text{I}$ (onset of infectiousness), and $\text{I}\rightarrow\text{R}$ (removal) transitions in the population between time $0$ and time $T$.  For all ordered pairs $ij$ such that $i$ is infected, we observe $C_{ij}$.  

\subsection{Censoring}
We assume that we can observe $\tau_{ij}$ only if $j$ is infected by $i$ at time $t_i + \varepsilon_i + \tau_{ij}$. Clearly, $\tau_{ij}$ can be observed only if $C_{ij} = 1$. The following processes can right-censor $\tau_{ij}$
\begin{enumerate}
    \item $\mathcal{I}_i(\tau) = \indicator{\tau \in (0, \iota_i]}$ indicates whether $i$ remains infectious at infectiousness age $\tau$. Thus, $i$ makes infectious contact with $j$ at infectiousness age $\tau_{ij}$ only if $\mathcal{I}_i(\tau_{ij}) = 1$.
    \item $\mathcal{S}_{ij}(\tau) = \indicator{t_i + \varepsilon_i + \tau \leq t_j}$ indicates whether $j$ remains susceptible when $i$ reaches infectiousness age $\tau$. Thus, $j$ can be infected by $i$ at time $t_{ij}$ only if $\mathcal{S}_{ij}(\tau_{ij}) = 1$.
    \item Assume that infection in person $j$ can be observed until time $T_j$. Then $\mathcal{Y}_{ij}(\tau) = \indicator{t_i + \varepsilon_i + \tau \leq T_j}$ indicates whether infection in $j$ can be observed when $i$ reaches infectiousness age $\tau$, so infectious contact from $i$ to $j$ can be observed at time $t_{ij}$ only if $\mathcal{Y}_{ij}(\tau_{ij}) = 1$.
\end{enumerate}
Since $\mathcal{I}_i(\tau)$, $\mathcal{S}_{ij}(\tau)$, and $\mathcal{Y}_{ij}(\tau)$ are left-continuous,
\begin{equation}
    Y_{ij}(\tau) = C_{ij} \mathcal{I}_i(\tau) \mathcal{S}_{ij}(\tau) \mathcal{Y}_{ij}(\tau)
\end{equation}
is a left-continuous process that indicates the risk of an observed infectious contact from $i$ to $j$ when $i$ reaches infectiousness age $\tau$. 

The assumptions made in the stochastic S(E)IR model above ensure that $\mathcal{I}_i(\tau)$ and $\mathcal{S}_{ij}(\tau)$ independently censor $\tau_{ij}$. The methods in this paper also assume that $\mathcal{Y}_{ij}(\tau)$ independently censors $\tau_{ij}$. When who-infects-whom is observed, $\mathcal{T}_{ij}$ can be any stopping time with respect to the history generated by $\mathcal{I}_i(\tau)$, $\mathcal{S}_{ij}(\tau)$, and other processes that independently censor $\tau_{ij}$. Our assumptions can be relaxed as long as independent censoring of $\tau_{ij}$ is preserved~\citep{Kenah5}. When who-infected-whom is not observed, we require that $\mathcal{Y}_{ij}(t_j - t_i - \varepsilon_i) = 1$ for all $i \in \mathcal{V}_j$ for each non-imported infection $j$. 

\subsection{Nonparametric survival analysis of epidemic data}
\label{sec:CIestimation}
Assume there is a hazard function $\lambda(\tau)$ such that $\lambda_{ij}(\tau) = \lambda(\tau)$ for each $ij$ such that $C_{ij} = 1$. Let $\Lambda(\tau) = \int_0^\tau \lambda(u) \dif u$ be the corresponding cumulative hazard function. \citet{Kenah5} extended the Nelson-Aalen estimator to obtain a nonparametric marginal Nelson-Aalen estimator of $\Lambda(\tau)$. This derivation used counting processes and martingales defined in infectiousness age.

For each ordered pair $ij$, let $\mathcal{N}_{ij}(\tau) = \indicator{\tau_{ij}\leq \tau}$ indicate whether infectious contact from $i$ to $j$ occurs by infectiousness age $\tau$ in person $i$, with $N_{ij}(\tau) = 0$ for all $\tau$ if $i$ is never infected. Then $\mathcal{N}_{ij}$ is continuous from the right with left-hand limits (cadlag) and $\mathcal{N}_{ij}(0) = 0$, so
\begin{equation}
    \mathcal{M}_{ij}(\tau) = \mathcal{N}_{ij}(\tau) - \int_0^\tau C_{ij} \Inf_i(u) \lambda(u) \dif u
    \label{eq:calMij}
\end{equation}
is a mean-zero martingale. Since we can observe infectious contact from $i$ to $j$ only if $j$ is susceptible and under observation, 
\begin{equation}
    N_{ij}(\tau) = \int_0^\tau Y_{ij}(u) \dif \mathcal{N}_{ij}(u),
\end{equation}
counts observed infectious contacts from $i$ to $j$ and
\begin{equation}
    M_{ij}(\tau) = N_{ij}(\tau) - \int_0^\tau Y_{ij}(u) \lambda(u) \dif u 
    \label{eq:initMij}
\end{equation}
is a mean-zero martingale.

\subsubsection{Who-infects-whom is observed}
The number of contact intervals of length $\geq \tau$ that were observed is 
\begin{equation}
    Y(\tau) = \sum_{j = 1}^n \sum_{i\neq j} Y_{ij}(\tau),
    \label{eq:Y}
\end{equation}
which is decreasing and left-continuous.  When who-infects-whom is observed, we can calculate the Nelson-Aalen estimator
\begin{equation}
    \hat{\Lambda}(\tau) = \int_0^\tau \frac{\indicator{Y(u) > 0}}{Y(u)}\dif N(u),
    \label{eq:hatLambda}
\end{equation}
where $N(\tau) = \sum_{j=1}^n\sum_{i\neq j} N_{ij}(\tau)$. For all $\tau$ such that $Y(\tau) > 0$, this is an unbiased estimator of $\Lambda(\tau)$ because
\begin{equation}
    \hat{\Lambda}(\tau) - \Lambda(\tau) = \int_0^\tau \frac{\indicator{Y(u) > 0}}{Y(u)}\dif M(u),
    \label{eq:hatLambdaM}
\end{equation}
where $M(\tau) = \sum_{j=1}^n\sum_{i\neq j} M_{ij}(\tau)$ is a mean-zero martingale.  When the contact interval distribution is continuous, the variance of $\hat{\Lambda}(\tau) - \Lambda(\tau)$ can be estimated using its optional variation process
\begin{equation}
    \hat{\sigma}^2(\tau) = \int_0^\tau \frac{1}{Y(u)^2}\dif N(u).
    \label{eq:hatsigma}
\end{equation}

\subsubsection{Who-infects-whom is not observed}
When who-infected-whom is not observed, we cannot calculate the Nelson-Aalen estimate because we do not know which contact intervals are censored and which are observed. Fortunately, 
\begin{equation}
    \Lambda(\tau) = \E[\big]{\hat{\Lambda}(\tau)} = \E[\big]{\condE{\hat{\Lambda}(\tau)}{\text{observed data}}}
\end{equation}
by the law of iterated expectation, so we can estimate $\Lambda(\tau)$ by estimating the mean of the possible Nelson-Aalen estimates. When the contact interval distribution is continuous, the probability that $j$ was infected by person $i$ given the observed history up to time $t_j$ is
\begin{equation}
    p_{ij}(\lambda) = \frac{\lambda_{ij}(t_j - t_i - \varepsilon_i) \indicator{i \in \mathcal{V}_j}}{\sum_{k\in\mathcal{V}_j}\lambda_{kj}(t_j - t_k - \varepsilon_k)},
    \label{eq:pij}
\end{equation}
Since the infector of each infected $j$ can be chosen independently given the observed data~\citep{Kenah3}, the probability of $\mathbf{v} = (v_1, \ldots, v_n)$ is
\begin{equation}
    \Pr(\mathbf{v}|\text{observed data}) = \prod_{j:\, 0 < v_j < \infty} p_{v_j j}.    
\end{equation}
Let $\hat{\Lambda}_\mathbf{v}(\tau)$ denote the $\hat{\Lambda}(\tau)$ that we would have calculated had we observed the transmission network $\mathbf{v}$. Then 
\begin{equation}
    \widetilde{\Lambda}(\tau) = \sum_{\mathbf{v}\in\mathcal{V}} \hat{\Lambda}_\mathbf{v}(\tau)\Pr(\mathbf{v}|\text{observed data})
    \label{eq:mNA}
\end{equation}
is an unbiased estimate of $\Lambda(\tau)$ for all $\tau$ such that $Y(\tau) > 0$.  We call this the marginal Nelson-Aalen estimate.  

Since the true hazard function $\lambda(\tau)$ is unknown, we cannot use equation~\eqref{eq:mNA} directly. Instead, we use it as part of an EM algorithm that starts from an initial guess at the hazard function. Given a hazard function $\lambda^{(k)}(\tau)$, let
\begin{equation}
    \widetilde{N}_{\cdot j}\big(\tau \big| \lambda^{(k)}\big) = \sum_{i\in\mathcal{V}_j} p_{ij}\big(\lambda^{(k)}\big)\indicator{\tau\geq t_j - t_i - \varepsilon_i}.
\end{equation}
Then the marginal Nelson-Aalen estimate given $\lambda^{(k)}$ is
\begin{equation}
    \widetilde{\Lambda}\big(\tau \big| \lambda^{(k)}\big) = \int_0^\tau \frac{1}{Y(u)} \dif \widetilde{N}\big(u \big| \lambda^{(k)}\big),
    \label{eq:tildeLambda}
\end{equation}
where $\widetilde{N}\big(\tau \big| \lambda^{(k)}\big) = \sum_{j=1}^n \widetilde{N}_{\cdot j}\big(\tau \big|\lambda^{(k)}\big)$. We can smooth the increments of $\widetilde{\Lambda}\big(\tau \big| \lambda^{(k)}\big)$ to estimate a new hazard function $\lambda^{(k+1)}$, and so on. Iterating from an initial $\lambda^{(0)}(\tau)$ leads to Algorithm~\ref{alg:EM}, which turns out to be EM algorithm. The limit of the sequence $\Lambda^{(k)}(\tau)$ is the marginal Nelson-Aalen estimate $\widetilde{\Lambda}(\tau)$~\citep{Kenah5}.
\begin{algorithm}
    Choose an initial $\lambda^{(0)}(\tau)$\;
    Set $k = 0$\;
    \While{convergence criterion not met}{
        \emph{E-step:} Calculate infector probabilities $p_{ij}\big(\lambda^{(k)}\big)$\;
        \emph{M-step:} Calculate $\Lambda^{(k)}(\tau) = \widetilde{\Lambda}\big(\tau\big|\lambda^{(k)}\big)$\;
        \emph{Smoothing step:} Smooth $\Lambda^{(k)}(\tau)$ to obtain $\lambda^{(k+1)}(\tau)$\;
        Set $k = k+1$\;
    }
    \caption{EM algorithm for nonparametric estimation of $\Lambda(\tau)$ using data from a homogeneous population.}
    \label{alg:EM}
\end{algorithm}

The variance of $\widetilde{\Lambda}(\tau)$ can be estimated using the conditional variance formula.  Conditioning on the transmission network $\mathbf{v}$, we have
\begin{equation}
    \widetilde{\sigma}^2(\tau) = \E[\big]{\hat{\sigma}^2_\mathbf{v}(\tau)} + \Var[\big]{\hat{\Lambda}_\mathbf{v}(\tau)},    
\end{equation}
where $\hat{\Lambda}_\mathbf{v}(\tau)$ is the Nelson-Aalen estimate from equation~\eqref{eq:hatLambda} and $\hat{\sigma}^2_\mathbf{v}(\tau)$ is the variance estimate from equation~\eqref{eq:hatsigma} that we would have calculated had we observed the transmission network $\mathbf{v}$.  This reduces to~\citep{Kenah5}
\begin{equation}
    \widetilde{\sigma}^2(\tau) = 2\int_0^\tau \frac{\indicator{Y(u)>0}}{Y(u)^2}\dif\widetilde{N}(u)\; - \sum_{j: t_j\leq T} \Big(\int_0^\tau \frac{\indicator{Y(u)>0}}{Y(u)}\dif\widetilde{N}_{\cdot j}(u)\Big)^2.
\end{equation}

\section{Methods}
\label{sec:methods}
The methods of Section~\ref{sec:CIestimation} assume a homogeneous population in the sense that $\lambda_{ij}(\tau)$ is the same for all $ij$ such that $C_{ij} = 1$.  Now consider a semiparametric relative-risk model like that of~\citet{PrenticeSelf} in which 
\begin{equation}
    \lambda_{ij}(\tau) = r\big(\beta_0^\trans X_{ij}(\tau)\big)\lambda_0(\tau),
\end{equation}
where $\lambda_0(\tau)$ is an unspecified baseline hazard function, $r:\mathbb{R} \rightarrow (0, \infty)$ is a relative risk function, $\beta_0$ is an unknown $b \times 1$ coefficient vector, and $X_{ij}(\tau)$ is a $b \times 1$ predictable covariate process taking values in a set $\mathcal{X}$. We assume that $r$ has continuous first and second derivatives, $r(0) = 1$, and $\ln r(\beta^\trans X)$ is bounded on $\mathcal{X}$. Letting $r(x) = \exp(x)$ gives us a loglinear relative risk regression model like that of~\citet{Cox1972}, and letting $r(x) = 1 + x$ gives us a linear relative risk regression model. 

To fit these semiparametric models, we adapt the nonparametric estimators from~\citet{Kenah5} to account for the relative risk function. First, we consider the case where who-infects-whom is observed. Then we describe an EM algorithm to handle the case where who-infects-whom is not observed. 

\subsection{Who-infects-whom is observed}
\label{sec:hat}
Let $\Lambda_0(\tau) = \int_0^\tau \lambda_0(u) \dif u$. For a given $\beta$, the Breslow estimator~\citep{Breslow} of $\Lambda_0(\tau)$ is
\begin{equation}
    \hat{\Lambda}(\beta, \tau) = \int_0^\tau \frac{1}{Y(\beta, u)}\dif N(u),
    \label{eq:Breslow}
\end{equation}
where
\begin{equation}
    Y(\beta, u) = \sum_{j=1}^n\sum_{i\neq j} r\big(\beta^\trans X_{ij}(u)\big)Y_{ij}(u).
\end{equation}
This estimator has two desirable properties.  First, $\hat{\Lambda}(\beta_0, \tau)$ is an unbiased estimator of $\Lambda_0(\tau)$.  For all $\tau$ such that $Y(\tau) > 0$,
\begin{equation}
    \hat{\Lambda}(\beta_0, \tau) - \Lambda_0(\tau) = \int_0^\tau \frac{\indicator{Y(u) > 0}}{Y(\beta_0, u)}\dif M(\beta_0, u),
\end{equation}
where 
\begin{equation}
    M(\beta_0, \tau) = N(\tau) - \int_0^\tau Y(\beta_0, u)\lambda_0(u) \dif u.
\end{equation}
is a mean-zero martingale. Second, $\hat{\Lambda}(\beta, \tau)$ maximizes the log likelihood
\begin{equation}
    \ell(\beta,\Lambda) = \sum_{j=1}^n \sum_{i \neq j} \ln\Big(r\big(\beta^\trans X_{v_jj}(\tau_{v_j j})\big) \dif \Lambda(\tau_{v_j j})\Big) - \int_0^\infty Y(\beta, u) \dif \Lambda(u),
    \label{eq:lbeta}
\end{equation}
over all step functions $\Lambda(\tau)$. Substituting $\hat{\Lambda}(\beta, \tau)$ into $\ell(\beta, \Lambda)$, we get the log profile likelihood
\begin{equation}
    \ell\big(\beta, \hat{\Lambda}\big) = \Bigg(\sum_{j=1}^n \ln\frac{r\big(\beta^\trans X_{v_jj}(\tau_{v_j j})\big)}{Y(\beta, \tau_{v_j j})}\Bigg) - \mathcal{T},
    \label{eq:plT}
\end{equation}
where $\mathcal{T} = \max\{\tau:Y(\tau) > 0\}$. The first term is similar to the log partial likelihood from~\citet{Cox1972} and the second term does not depend on $\beta$. Dropping the second term, let
\begin{equation}
    pl(\beta) = \sum_{j=1}^n \ln\frac{r\big(\beta^\trans X_{v_jj}(\tau_{v_j j})\big)}{Y(\beta, \tau_{v_j j})}
\end{equation}
be the log partial likelihood for $\beta$. This derivation of the partial likelihood as a profile likelihood follows that of~\citet{Johansen1983}. Let $\hat{\beta}$ denote the value of $\beta$ that maximizes $pl(\beta)$, and let $\hat{\Lambda}_0(\tau) = \hat{\Lambda}(\hat{\beta}, \tau)$ denote the corresponding Breslow estimate of the baseline cumulative hazard.

\subsubsection{Partial likelihood score process}
We can rewrite $pl(\beta)$ as a sum of stochastic integrals:
\begin{equation}
    pl(\beta) = \sum_{j=1}^n\sum_{i\neq j} \int_0^\infty \ln \frac{r\big(\beta^\trans X_{ij}(u)\big)}{Y(\beta, u)}\dif N_{ij}(u).
    \label{eq:pl}
\end{equation}
The corresponding score process is
\begin{equation}
    U(\beta, \tau) = \sum_{j=1}^n \sum_{i\neq j} \int_0^\tau \deriv{\beta} \ln r\big(\beta^\trans X_{ij}(u)\big) - E(\beta, u) \dif N_{ij}(u),
    \label{eq:Ubeta}
\end{equation}
where 
\begin{equation}
    E(\beta, u) = \frac{\sum_{j=1}^n \sum_{i\neq j} r\big(\beta^\trans X_{ij}(u)\big) Y_{ij}(u) \deriv{\beta} \ln r\big(\beta^\trans X_{ij}(u)\big)}{\sum_{j=1}^n \sum_{i\neq j} r\big(\beta^\trans X_{ij}(u)\big) Y_{ij}(u)}.
\end{equation}
is the expected value of $\deriv{\beta} \ln r\big(\beta^\trans X_{ij}(u)\big)$ over the risk set at $u$ when each pair is weighted by its hazard of transmission at $u$. By the Doob-Meyer decomposition, there is a mean-zero martingale $M_{ij}(u)$ for each $ij$ such that
\begin{equation}
    \dif N_{ij}(u) = r\big(\beta_0^\trans X_{ij}(u)\big) \lambda_0(u) Y_{ij}(u) \dif u + \dif M_{ij}(u).
    \label{eq:DoobMeyer}
\end{equation}
Expanding equation~\eqref{eq:Ubeta} using this decomposition and simplifying, we get
\begin{equation}
    U(\beta_0, \tau) = \sum_{j=1}^n \sum_{i\neq j} \int_0^\tau \deriv{\beta} \ln \frac{r\big(\beta_0^\trans X_{ij}(u)\big)}{Y(\beta_0, u)} \dif M_{ij}(u).
\end{equation}
Since it is a sum of integrals of predictable processes with respect to martingales, $U(\beta_0, \tau)$ is a mean-zero martingale. 

\subsubsection{Observed and expected information}
Since the $N_{ij}(\tau)$ do not jump simultaneously in continuous time, the predictable variation process of $U(\beta_0, \tau)$ is
\begin{equation}
    \langle U(\beta_0)\rangle(\tau) = \int_0^\tau V(\beta_0, u) Y(\beta_0, u) \lambda_0(u) \dif u,
    \label{eq:angleU}
\end{equation}
where 
\begin{equation}
    V(\beta, u) = \sum_{j=1}^n\sum_{i\neq j} \bigg(\deriv{\beta}\ln \frac{r\big(\beta^\trans X_{ij}(u)\big)}{Y(\beta, u)}\bigg)^{\otimes 2} \frac{r\big(\beta^\trans X_{ij}(u)\big) Y_{ij}(u)}{Y(\beta, u)}
\end{equation}
is the variance of $\deriv{\beta} \ln r\big(\beta^\trans X_{ij}(u)\big)$ over the risk set at $u$ when each pair $ij$ is weighted by its hazard of transmission at $u$. 

Let $I(\beta) = -\secderiv{\beta} pl(\beta)$ be the observed information. Then
\begin{align}
    I(\beta) &= \sum_{j=1}^n \sum_{i \neq j} \int_0^\infty \bigg(\deriv{\beta} \ln r\big(\beta^\trans X_{ij}(u)\big)\bigg)^{\otimes 2} - E(\beta, u)^{\otimes 2} \dif N_{ij}(u) \nonumber \\
     &\qquad - \sum_{j=1}^n \sum_{i\neq j} \int_0^\infty \frac{\secderiv{\beta} r\big(\beta^\trans X_{ij}(u)\big)}{r\big(\beta^\trans X_{ij}(u)\big)} - \frac{\secderiv{\beta} Y(\beta, u)}{Y(\beta, u)} \dif N_{ij}(u),
     \label{eq:Iobserved}
\end{align}
where $v^{\otimes 2} = vv^\trans$ for a column vector $v$ ($v^2$ for scalar $v$). Expanding $I(\beta_0)$ via the Doob-Meyer decomposition~\eqref{eq:DoobMeyer} and simplifying, we get
\begin{align}
    I(\beta_0) 
    &= \int_0^\infty V(\beta_0, u) Y(\beta_0, u) \lambda_0(u) \dif u \nonumber \\
    &\qquad + \sum_{j=1}^n \sum_{i \neq j} \int_0^\infty \secderiv{\beta} \ln \frac{r\big(\beta_0^\trans X_{ij}(u)\big)}{Y(\beta_0, u)} \dif M_{ij}(u).
\end{align}
The second term has expectation zero, so $I(\beta_0)$ is an unbiased estimate of the variance of $U(\beta_0, \infty)$.

Another estimate of $\Var[\big]{U(\beta_0, \infty)}$ is obtained by substituting the increments of the Breslow estimator~\eqref{eq:Breslow} for $\lambda_0(u) \dif u$ in equation~\eqref{eq:angleU}. This gives us the (estimated) expected information 
\begin{equation}
    \mathcal{I}(\beta) = \int_0^\infty V(\beta, u) \dif N(u).
    \label{eq:Iexpected}
\end{equation}
Expanding $\mathcal{I}(\beta_0)$ using the Doob-Meyer decomposition and simplifying, we get
\begin{equation}
    \mathcal{I}(\beta_0) = \int_0^\infty V(\beta_0, u) Y(\beta_0, u) \lambda_0(u) \dif u + \int_0^\infty V(\beta_0, u) \dif M(u).
\end{equation}
The second term has expectation zero, so $\mathcal{I}(\beta_0)$ is also an unbiased estimate of the variance of $U(\beta_0, \infty)$. $\mathcal{I}(\beta_0)$ may be a better estimator of $\Var[\big]{U(\beta_0, \infty)}$ than $I(\beta_0)$ because it is guaranteed to be positive semidefinite~\citep{PrenticeSelf} and it depends only on aggregates over risk sets~\citep{Aalen}. 

When $r(x) = e^x$ as in the Cox model, $\secderiv{\beta} \ln r\big(\beta^\trans X\big) = 0$ for all $\beta$ and $X$ so
\begin{equation}
    I(\beta) = \int_0^\infty \secderiv{\beta} \ln Y(\beta, u) \dif N(u).
\end{equation}
Since
\begin{equation}
    \secderiv{\beta} r\big(\beta^\trans X\big) = \Big(\deriv{\beta} \ln r\big(\beta^\trans X\big)\Big)^{\otimes 2} r\big(\beta^\trans X\big),
\end{equation}
for all $\beta$ and $X$, we have $\secderiv{\beta} \ln Y(\beta, u) = V(\beta, u)$. Therefore, $I(\beta) = \mathcal{I}(\beta)$ for all $\beta$. For general $r\big(\beta^\trans X\big)$, $I(\beta_0)$ and $\mathcal{I}(\beta_0)$ are asymptotically equivalent under weak regularity conditions (see Appendix~\ref{app:asymptotics}).

\subsubsection{Large-sample estimation of $\beta_0$ and $\Lambda_0(\tau)$}
Appendix~\ref{app:asymptotics} outlines sufficient conditions for the asymptotic normality of $U(\beta_0, \tau)$ and $\hat{\beta}$ as $m \rightarrow \infty$, where $m$ is the number of pairs $ij$ at risk of transmission. These are the same conditions required for asymptotic normality in standard survival data, except for the requirement that $m$ is much larger than the largest number of infectors to which any single susceptible is exposed. Under these conditions, hypothesis tests and confidence intervals for $\beta_0$ can be obtained using score, Wald, or likelihood ratio statistics. 

Given $\hat{\beta}$, the Breslow estimator of $\Lambda_0(\tau)$ is $\hat{\Lambda}_0(\tau) = \hat{\Lambda}_0(\hat{\beta}, \tau)$. Its variance is consistently estimated by 
\begin{equation}
    \hat{\sigma}^2_0(\tau) = \bigg(\deriv{\beta} \hat{\Lambda}(\hat{\beta}, \tau)\bigg)^\trans I(\hat{\beta})^{-1} \bigg(\deriv{\beta} \hat{\Lambda}(\hat{\beta}, \tau)\bigg) + \int_0^\tau \frac{1}{Y(\hat{\beta}, u)^2} \dif N(u),
    \label{eq:hatsigma}
\end{equation}
which is derived in Appendix~\ref{app:hatVariance}. $I(\hat{\beta})$ can be replaced by $\mathcal{I}(\hat{\beta})$. Using the martingale central limit theorem and a log transformation, we get the approximate pointwise $1 - \alpha$ confidence limits
\begin{equation}
    \hat{\Lambda}_0(\tau) \exp\bigg(\pm \frac{\hat{\sigma}_0(\tau)}{\hat{\Lambda}_0(\tau)} \Phi^{-1}\Big(1 - \frac{\alpha}{2}\Big)\bigg).
\end{equation}
Point and interval estimates for the baseline survival function can be obtained using the product integral~\citep{Aalen} or using $S_0(\tau) = \exp\big(-\Lambda_0(\tau)\big)$. These estimates are asymptotically equivalent, but the latter is more consistent with the derivation of the partial likelihood as a profile likelihood.

\subsection{Who-infects-whom is not observed}
\label{sec:tilde}
If we observe infection but not who-infects-whom, we cannot calculate the partial likelihood $pl(\beta)$ or the Breslow estimate $\hat{\Lambda}(\beta, \tau)$ because we do not know which contact intervals are observed and which are censored. However, we can use an EM algorithm similar to that of~\citet{Kenah5} to obtain consistent and asymptotically normal estimates of $\beta_0$ and $\Lambda_0(\tau)$.

Given a coefficient vector $\beta$ and a baseline hazard function $\lambda(\tau)$, we can calculate $\Pr(\mathbf{v}|\text{observed data})$ for each $\mathbf{v} \in \mathcal{V}$~\citep{Kenah3}. If $j$ is infected at time $t_j$, the probability that $j$ was infected by $i$ given $\beta$ and $\lambda(\tau)$ is
\begin{equation}
    p_{ij}(\beta, \lambda) = \frac{r\big(\beta^\trans X_{ij}(t_j - t_i - \varepsilon_i)\big) \lambda(t_j - t_i - \varepsilon_i) \indicator{i\in\mathcal{V}_j}}{\sum_{k \in \mathcal{V}_j} r\big(\beta^\trans X_{kj}(t_j - t_k - \varepsilon_k)\big) \lambda(t_j - t_k - \varepsilon_k)}.
\end{equation}
The infectors of different infected persons can be chosen independently, so the probability of a transmission network $\mathbf{v} = (v_1, \ldots, v_n)$ given $\beta$, $\lambda(\tau)$, and the observed data is
\begin{equation}
    \Pr(\mathbf{v}|\beta, \lambda, \text{observed data}) = \prod_{j:\, 0 < v_j < \infty} p_{v_j j}(\beta, \lambda).
\end{equation}
Note that these equations assume a continuous contact interval distribution, so simultaneous infectious contacts have probability zero.

Let $pl_\mathbf{v}(\beta)$ be the log partial likelihood that we would have calculated had we observed the transmission network $\mathbf{v}$. Given a coefficient vector $\beta^*$ and a baseline hazard function $\lambda^*(\tau)$, the expected log likelihood is
\begin{align}
    \widetilde{pl}_{\beta^*, \lambda^*}(\beta) 
    &= \sum_{\mathbf{v} \in \mathcal{V}} pl_\mathbf{v}(\beta) \Pr(\mathbf{v} | \beta^*, \lambda^*, \text{ observed data}) \nonumber \\
    &= \sum_{j=1}^n \sum_{i\neq j} \int_0^\mathcal{T} \ln \frac{r\big(\beta^\trans X_{ij}(u)\big)}{Y(\beta, u)} \dif \widetilde{N}_{ij}(u | \beta^*, \lambda^*),
\end{align}
where $\widetilde{N}_{ij}(\tau | \beta^*, \lambda^*) = p_{ij}(\beta^*, \lambda^*) \indicator{\tau \geq t_j - t_i - \varepsilon_i}$. Now let $N(\tau|\mathbf{v})$ be the value of $N(\tau)$ that we would have calculated had we observed the transmission network $\mathbf{v}$ and let the corresponding Breslow estimate be
\begin{equation}
    \hat{\Lambda}_\mathbf{v}(\beta, \tau) = \int_0^\tau \frac{1}{Y(\beta, u)} \dif N(u|\mathbf{v}).
\end{equation}
Then the the marginal Breslow estimate given $\beta^*$ and $\lambda^*(\tau)$ is
\begin{align}
    \widetilde{\Lambda}_{\beta^*, \lambda^*}(\beta, \tau) 
    &= \sum_{\mathbf{v} \in \mathcal{V}} \hat{\Lambda}_{\mathbf{v}}(\beta, \tau) \Pr(\mathbf{v}|\beta^*, \lambda^*, \text{observed data}) \nonumber \\
    &= \int_0^\tau \frac{1}{Y(\beta, u)} \dif \widetilde{N}(u | \beta^*, \lambda^*),
\end{align}
where $\widetilde{N}(\tau |\beta^*, \lambda^*) = \sum_{j=1}^n \sum_{i \neq j} \widetilde{N}_{ij}(\tau |\beta^*, \lambda^*)$.

For the relative risk function $r(x) = e^x$, the expected log partial likelihood $\widetilde{pl}_{\beta^*, \lambda^*}(\beta)$ is the log partial likelihood of a weighted Cox regression model~\citep{TherneauGrambsch} with two copies of each pair $ij$: an uncensored copy with weight $p_{ij}(\beta^*, \lambda^*)$ and a censored copy with weight $1 - p_{ij}(\beta^*, \lambda^*)$. The baseline hazard estimate from this model is the marginal Breslow estimate $\widetilde{\Lambda}_{\beta^*, \lambda^*}(\widetilde{\beta}, \tau)$, where $\widetilde{\beta} = \arg\max_{\beta} \widetilde{pl}_{\beta^*, \lambda^*}(\beta)$.

\subsubsection{EM algorithm}
When who-infects-whom is not observed, the semiparametric regression model can be fit using the ECM algorithm of~\citet{MengRubin1993}, which is an extension of the EM algorithm of~\citet{DempsterLairdRubin}. In each iteration, we first estimate $\beta_0$ using the expected log partial likelihood and then calculate the marginal Breslow estimator of $\Lambda_0(\tau)$. We then use these new estimates to re-weight the possible $\mathbf{v}$. The entire process is described in Algorithm~\ref{alg:ECM}.
\begin{algorithm}
    Choose an initial $\beta^{(0)}$ and $\lambda^{(0)}(\tau)$\;
    Set $k = 0$\;
    \While{convergence criterion not met}{
        \emph{E-step:} Calculate infector probabilities $p_{ij}\big(\beta^{(k)}, \lambda^{(k)}\big)$ \;
        \emph{CM1-step:} Find $\beta^{(k+1)} = \arg\max_\beta \widetilde{pl}_{\beta^{(k)}, \lambda^{(k)}}(\beta)$ \;
        \emph{CM2-step:} Calculate $\widetilde{\Lambda}^{(k+1)}_0(\tau) = \widetilde{\Lambda}_{\beta^{(k)}, \lambda^{(k)}}\big(\beta^{(k+1)}, \tau\big)$ \;
        \emph{Smoothing step:} Smooth $\Lambda^{(k+1)}(\tau)$ to obtain $\lambda^{(k+1)}(\tau)$\;
        Set $k = k+1$\;
    }
    \caption{ECM algorithm for semiparametric estimation of $\beta_0$ and $\Lambda_0(\tau)$ in a heterogeneous population.}
    \label{alg:ECM}
\end{algorithm}

To show that this is an ECM algorithm, we must show that the CM1 and CM2 steps are conditional maximizations of the expected log likelihood. Since the CM1 step is a conditional maximization by definition, it remains to show that the CM2 step is a conditional maximization. Given a coefficient vector $\beta^*$ and a hazard function $\lambda^*$, the expected log likelihood is
\begin{align}
    \widetilde{\ell}_{\beta^*, \lambda^*}(\beta, \Lambda) = & \sum_{j=1}^n \sum_{i\neq j} p_{ij}(\beta^*, \lambda^*) \ln \Big(r\big(\beta^\trans X_{ij}(t_j - t_i - \varepsilon_i)\big) \dif \Lambda(t_j - t_i - \varepsilon_i)\Big) \nonumber \\
    & \qquad -\int_0^\infty Y(\beta, u) \dif \Lambda(u).
\end{align}
Differentiating with respect to $\dif \Lambda(t_j - t_i - \varepsilon_i)$ for each $i$ and $j$ shows that, for a fixed $\beta$, $\widetilde{\ell}_{\beta^*, \lambda^*}(\beta, \Lambda)$ is maximized over all step functions $\Lambda(\tau)$ by setting
\begin{equation}
    \dif{\Lambda}(t_j - t_i - \varepsilon_i) = \frac{p_{ij}(\beta^*, \lambda^*)}{Y(\beta, t_j - t_i - \varepsilon_i)},
\end{equation}
exactly as in the marginal Breslow estimator $\widetilde{\Lambda}_{\beta^*, \lambda^*}(\beta, \tau)$. Therefore, Algorithm~\ref{alg:ECM} is an ECM algorithm. When it is known that $\beta = 0$, it reduces to Algorithm~\ref{alg:EM}, which shows that convergence of both $\beta^{(k)}$ and $\Lambda^{(k)}(\tau)$ should be monitored to ensure convergence of the ECM algorithm.

\subsubsection{Large-sample estimation of $\beta_0$}
Let $\widetilde{\beta}$ denote the estimate of $\beta_0$ to which the ECM algorithm converges, and let $\widetilde{\lambda}(\tau)$ denote the corresponding estimate of $\lambda_0(\tau)$. Let $U_\mathbf{v}(\tau, \beta)$ and $I_\mathbf{v}(\beta)$ denote the score and the observed information that we would have calculated had we observed the transmission network $\mathbf{v}$. Using the methods of~\citet{Louis1982}, the observed information is 
\begin{equation}
    \widetilde{I}(\widetilde{\beta}) = \subE[\big]{\widetilde{\beta}, \widetilde{\lambda}}{I_\mathbf{v}(\widetilde{\beta})} - \subE[\big]{\widetilde{\beta}, \widetilde{\lambda}}{U_\mathbf{v}(\widetilde{\beta}, \infty)^{\otimes 2}},
    \label{eq:tildeI}
\end{equation}
where $\subE{\beta, \lambda}{\cdot}$ denotes an expectation taken under the assumption that the true coefficient vector is $\beta$ and the true baseline hazard function is $\lambda(\tau)$. The first term in~\eqref{eq:tildeI} is 
\begin{equation}
    - \sum_{j=1}^n \sum_{i\neq j} \int_0^\tau \secderiv{\beta} \ln \frac{r\big(\widetilde{\beta}^\trans X_{ij}(u)\big)}{Y(\beta, u)} \dif \widetilde{N}_{ij}(u),
\end{equation}
where $\widetilde{N}_{ij}(u) = \widetilde{N}_{ij}(u | \widetilde{\beta}, \widetilde{\lambda})$. This is the observed information matrix from a weighted regression model where each $ij$ has an uncensored copy with weight $p_{ij}(\widetilde{\beta}, \widetilde{\lambda})$ and a censored copy with weight $1 - p_{ij}(\widetilde{\beta}, \widetilde{\lambda})$. To evaluate the second term in~\eqref{eq:tildeI}, let 
\begin{equation}
    \widetilde{U}_{\cdot j}(\beta, \tau) = \sum_{i\neq j} \int_0^\tau \deriv{\beta} \ln \frac{r\big(\beta^\trans X_{ij}(u)\big)}{Y(\beta, u)} \dif \widetilde{N}_{ij}(u),
\end{equation}
be the expected score contribution from individual $j$ as a susceptible. Then $\subE[\big]{\widetilde{\beta}, \widetilde{\lambda}}{U(\widetilde{\beta}, \infty)^{\otimes 2}}$ is
\begin{equation}
    \sum_{j=1}^n \sum_{i\neq j} \int_0^\infty \bigg(\deriv{\beta} \ln \frac{r\big(\widetilde{\beta}^\trans X_{ij}(u)\big)}{Y(\widetilde{\beta}, u)}\bigg)^{\otimes 2} \dif \widetilde{N}_{ij}(u) - \sum_{j=1}^n \widetilde{U}_{\cdot j}(\widetilde{\beta}, \infty)^{\otimes 2}
\end{equation}
because $\sum_{j=1}^n \widetilde{U}_{\cdot j}(\widetilde{\beta}, \infty) = 0$, each infected person $j$ has only one infector in any $\mathbf{v}$, and the infectors of different individuals can be chosen independently.

\subsubsection{Large-sample estimation of $\Lambda_0(\tau)$}
Let $\widetilde{\Lambda}_0(\tau)$ be the marginal Breslow estimate obtained after convergence of the ECM algorithm. Its variance is consistently estimated by 
\begin{align}
    \widetilde{\sigma}_0^2(\tau) 
    &=  \bigg(\deriv{\beta} \widetilde{\Lambda}_{\widetilde{\beta}, \widetilde{\lambda}}(\widetilde{\beta}, \tau)\bigg)^\trans \widetilde{I}(\widetilde{\beta})^{-1} \bigg(\deriv{\beta} \widetilde{\Lambda}_{\widetilde{\beta}, \widetilde{\lambda}}(\widetilde{\beta}, \tau)\bigg) \\
    &\qquad + 2 \int_0^\tau \frac{1}{Y(\widetilde{\beta}, u)^2} \dif \widetilde{N}(u) - \sum_{j=1}^n \bigg(\int_0^\tau \frac{1}{Y(\widetilde{\beta}, u)} \dif \widetilde{N}_{\cdot j}(u)\bigg)^2,
\end{align}
where $\widetilde{N}_{\cdot j}(u) = \sum_{i\neq j} \widetilde{N}_{ij}(u)$ (see Appendix~\ref{app:tildeVariance}). Using the martingale central limit theorem and a log transformation, we get the approximate pointwise $1 - \alpha$ confidence limits
\begin{equation}
    \widetilde{\Lambda}_0(\tau) \exp\bigg(\pm \frac{\widetilde{\sigma}_0(\tau)}{\widetilde{\Lambda}_0(\tau)} \Phi^{-1}\Big(1 - \frac{\alpha}{2}\Big)\bigg).
\end{equation}
As before, point and interval estimates for the baseline survival function can be obtained using the product integral~\citep{Aalen, Kenah5} or using $S_0(\tau) = \exp\big(-\Lambda_0(\tau)\big)$.

\section{Simulations}
\label{sec:sims}
The performance of the methods from section~\ref{sec:methods} was tested with a series of $12000$ network-based epidemic simulations. All epidemics took place on a Watts-Strogatz small-world network~\citep{WattsStrogatz}, which mimics the high clustering and low diameter of real human contact networks. Starting with a ring of $50000$ nodes, each node was connected to its $10$ nearest neighbors and each edge was rewired to a randomly chosen node with probability $0.1$. A new contact network was built for each simulation. 

All epidemic models were written in Python 2.7 (\texttt{www.python.org}) using the packages NetworkX 1.6 (\texttt{networkx.lanl.gov}), NumPy 1.6, and SciPy 0.9 (\texttt{www.scipy.org}). Statistical analysis was done in in R 2.15 (\texttt{www.r-project.org}) via the Rpy2 2.2 package (\texttt{rpy.sourceforge.net}). The code for the models is available as Online Supplementary Information.

\subsection{Transmission model}
The transmission model had a latent period of zero and an exponential infectious period with mean one. The baseline contact interval distribution was Weibull($\alpha$, $\gamma$), where $\alpha$ is the shape parameter and $\gamma$ is the rate parameter. $6000$ simulations had a Weibull($0.5$, $0.2$) distribution, which has $\Lambda_0(\tau) = (0.2 \tau)^{0.5}$. The other $6000$ had a Weibull($2$, $0.6$) distribution, which has $\Lambda_0(\tau) = (0.6 \tau)^2$. These distributions gave $R_0 \approx 3$ in a null model.

In the transmission model, each person $i$ had an infectiousness covariate $X_i^\text{inf}$ and a susceptibility covariate $X_i^\text{sus}$. Each pair $ij$ connected by an edge had a pairwise covariate $X_{ij}^\text{pair}$. All covariates were independent Bernoulli($.5$) random variables. For a connected pair $ij$, the hazard of transmission from $i$ to $j$ at infectiousness age $\tau$ of $i$ was
\begin{equation}
    \lambda_{ij}(\tau) = \exp\Big(\beta_\text{inf} X_i^\text{inf} + \beta_\text{sus} X_j^\text{sus} + \beta_\text{pair} X_{ij}^\text{pair}\Big)\lambda_0(\tau)
\end{equation}
For each parameter $\beta$, there were $4000$ simulations where its true value was chosen from a uniform distribution on $(-1, 1)$. Of these, $2000$ simulations used the Weibull($0.5$, $0.2$) baseline hazard and $2000$ used the Weibull($2$, $0.6$) baseline hazard. Of the $2000$ simulations for each baseline hazard, $1000$ had the other two $\beta$ set to $0$ and $1000$ had the other two $\beta$ set to $1$. 

Each simulated epidemic began with a single person infected at time $0$. Data from the next $1000$ infections was used to fit two regression models, one using information on who-infected-whom as in Section~\ref{sec:hat} and one using an EM algorithm as in Section~\ref{sec:tilde}. The EM algorithm used a minimum of $2$ and a maximum of $25$ iterations. At each iteration, a weighted Cox model was run using the last parameter estimates as the initial parameter estimate. Convergence was defined as a change less than $0.002$ in the expected log likelihood (tighter convergence criteria yielded nearly identical parameter estimates). After convergence, a Cox model was run using the final weights and initial parameters $\beta_\text{inf} = \beta_\text{sus} = \beta_\text{pair} = 0$. 

After each simulation, we recorded true values, estimates, and confidence intervals for each $\beta$ in the model and baseline hazard estimates and confidence intervals at the $10^\text{th}$, $25^\text{th}$, $50^\text{th}$, $75^\text{th}$, and $90^\text{th}$ percentiles of all possible (censored and uncensored) contact intervals. We also recorded the $\alpha$ and $\gamma$ of the baseline hazard function and the number of EM iterations.

\subsection{Results}
\begin{figure}[p]
    \includegraphics[width = \textwidth]{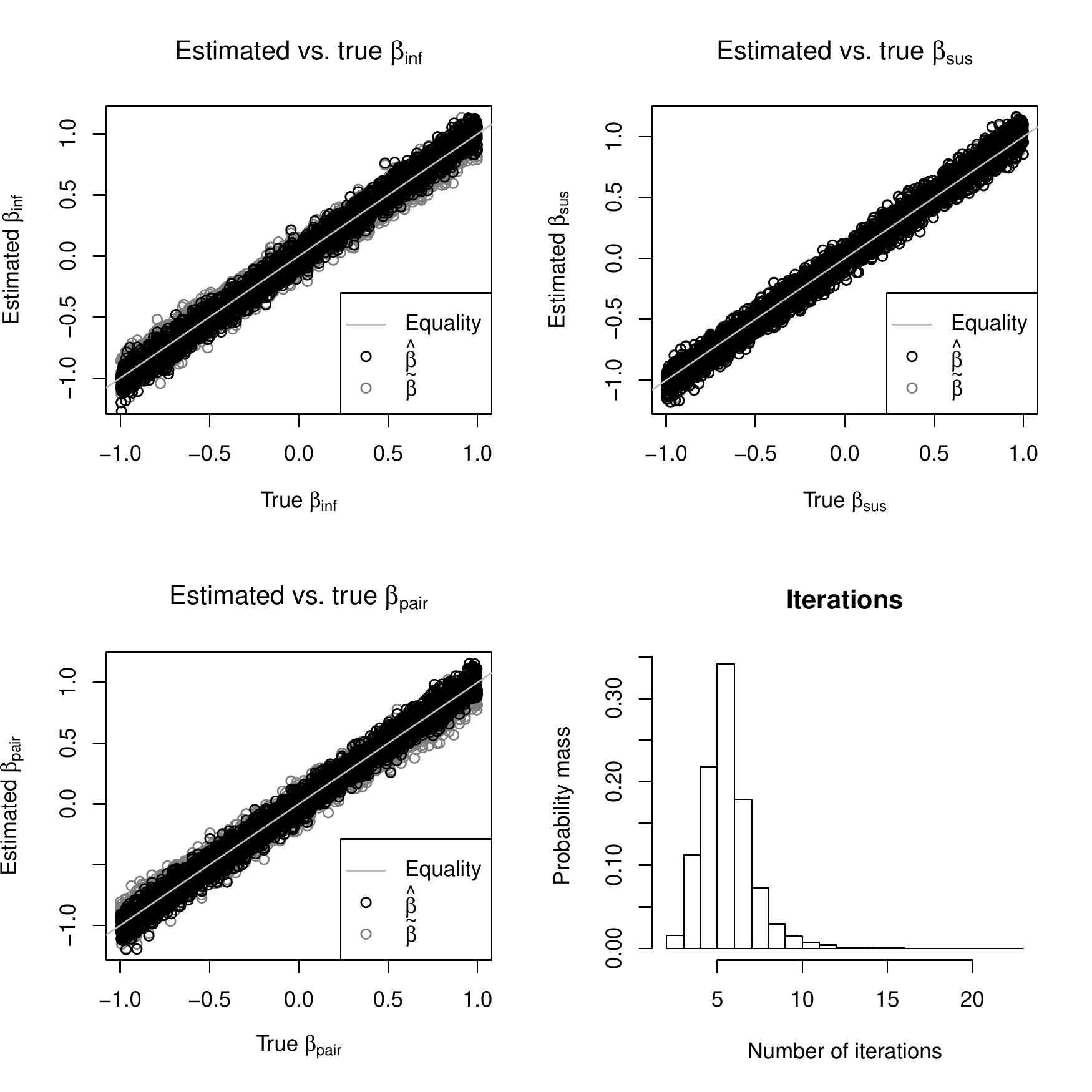}
    \caption{The top two panels and the bottom left panel show $\hat{\beta}$ (black circles) and $\widetilde{\beta}$ (gray circles) versus true $\beta$ for $\beta_\text{inf}$, $\beta_\text{sus}$, and $\beta_\text{pair}$. The bottom right panel shows a histogram of the number of EM iterations required for convergence.}
    \label{fig:beta}
\end{figure}

Figure~\ref{fig:beta} shows good agreement between the estimated and true $\beta_\text{inf}$, $\beta_\text{sus}$, and $\beta_\text{pair}$ for both $\hat{\beta}$ and $\widetilde{\beta}$. Table~\ref{tab:betaCI} shows 95\% confidence interval coverage probabilities above $.91$ for all combinations of baseline hazards and parameters. The lower right panel of Figure~\ref{fig:beta} shows that this was achieved with relatively few iterations. The median number of iterations was $6$, $98\%$ of simulations required $\leq 10$ iterations, and only $3$ out of $12000$ simulations failed to converge. 

\begin{figure}[p]
    \includegraphics[width = \textwidth]{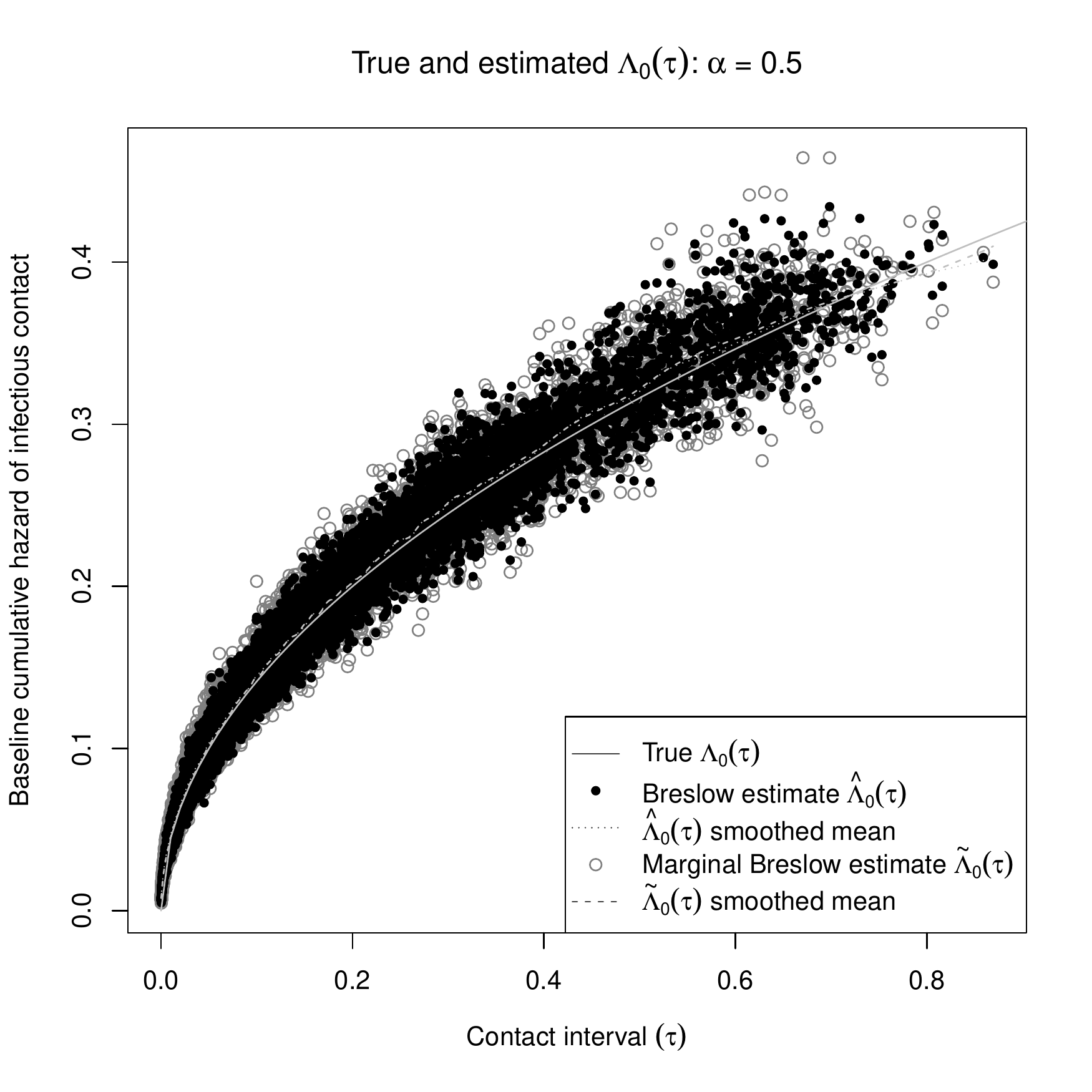}
    \caption{$\hat{\Lambda}_0(\tau)$ (black circles) and $\widetilde{\Lambda}_0(\tau)$ (gray circles) versus true $\Lambda_0(\tau)$ for the $6000$ simulations with a Weibull($0.5$, $0.2$) baseline contact interval distribution. For each simulation, a circle is shown for the $10^\text{th}$, $25^\text{th}$, $50^\text{th}$, $75^\text{th}$, and $90^\text{th}$ percentiles of all possible contact intervals. The smoothed means were calculated using cubic smoothing splines.}
    \label{fig:Lambda.5}
\end{figure}

\begin{figure}[p]
    \includegraphics[width = \textwidth]{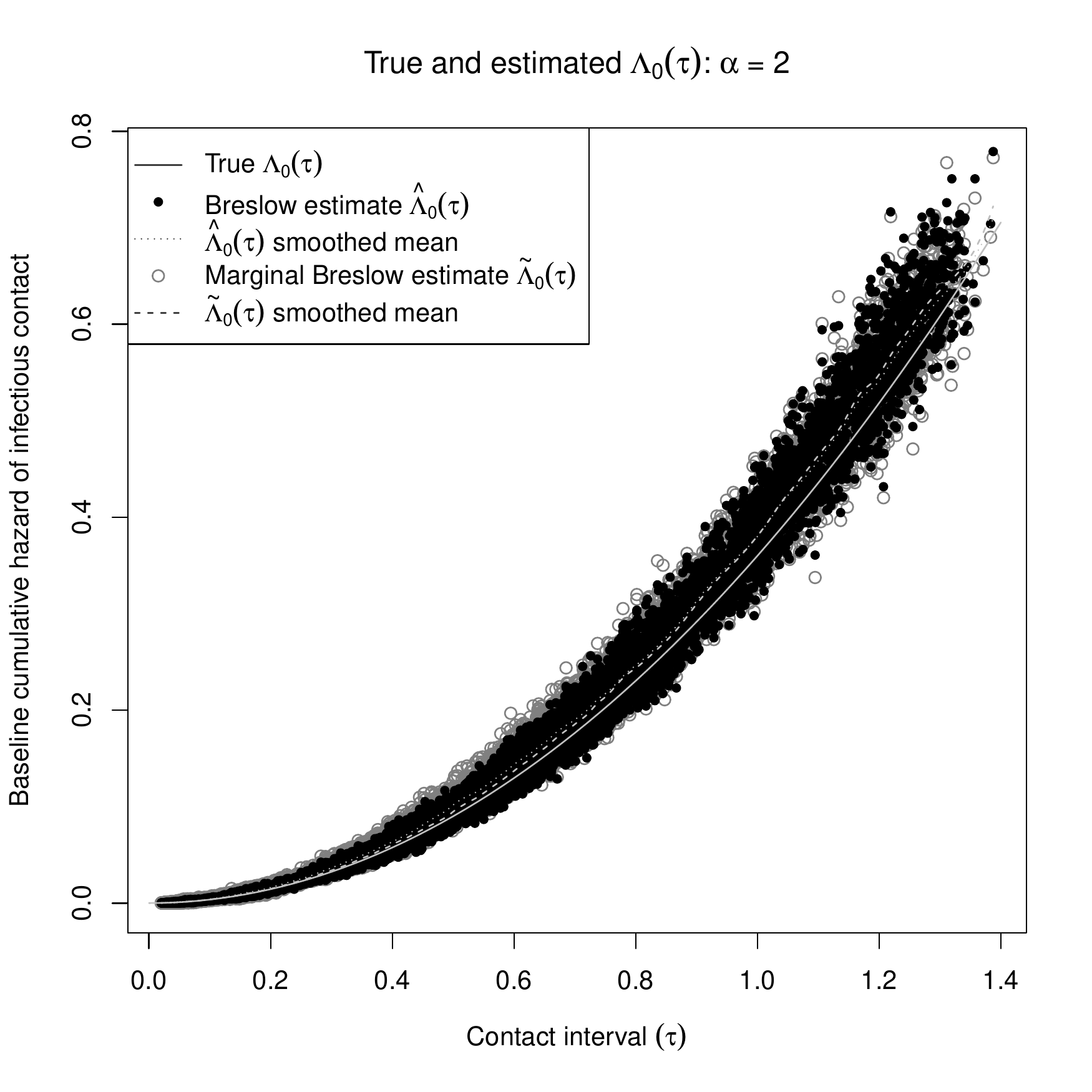}
    \caption{$\hat{\Lambda}_0(\tau)$ (black circles) and $\widetilde{\Lambda}_0(\tau)$ (gray circles) versus true $\Lambda_0(\tau)$ for the $6000$ simulations with a Weibull($2$, $0.6$) baseline contact interval distribution. For each simulation, a circle is shown for the $10^\text{th}$, $25^\text{th}$, $50^\text{th}$, $75^\text{th}$, and $90^\text{th}$ percentiles of all possible contact intervals. The smoothed means were calculated using cubic smoothing splines.}
    \label{fig:Lambda2}
\end{figure}

Figures~\ref{fig:Lambda.5} and~\ref{fig:Lambda2} show good agreement between the estimated and true baseline hazard for both $\hat{\Lambda}_0(\tau)$ and $\widetilde{\Lambda}_0(\tau)$. The smoothed means show almost no bias in $\hat{\Lambda}_0(\tau)$ or $\widetilde{\Lambda}_0(\tau)$ for $\alpha = .5$ and a slight upward bias at high $\tau$ for $\alpha = 2$. Table~\ref{tab:LambdaCI} shows good 95\% confidence interval coverage probabilites for the baseline hazard with shape parameter $\alpha = .5$ but much poorer coverage probabilities for the baseline hazard with $\alpha = 2$. When $\alpha = 2$, the baseline hazard function is changing fastest at high $\tau$, where there is the least data. Also, the estimated $\Lambda_0(\tau)$ and its confidence limits were evaluated as step functions; coverage probabilities may have been higher had smoothing or interpolation been used.

\begin{figure}[p]
    \includegraphics[width = \textwidth]{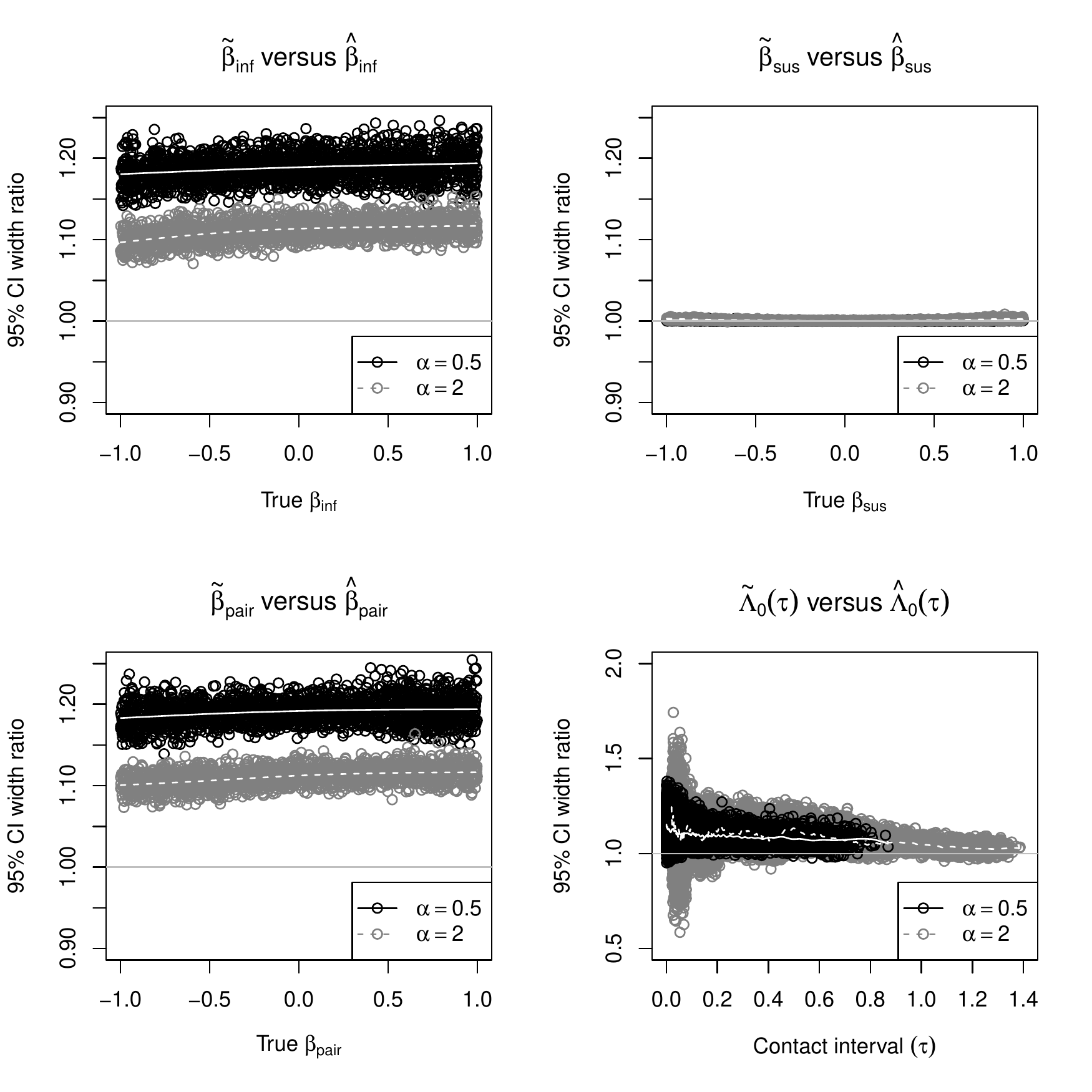}
    \caption{The width of $95\%$ confidence intervals for $\widetilde{\beta}_\text{inf}$, $\widetilde{\beta}_\text{sus}$, $\widetilde{\beta}_\text{pair}$, and $\widetilde{\Lambda}_0(\tau)$ in terms of the confidence interval width of $\hat{\beta}_\text{inf}$, $\hat{\beta}_\text{sus}$, $\hat{\beta}_\text{pair}$, and $\hat{\Lambda}_0(\tau)$. The solid gray lines show smoothed means for $\alpha = 0.5$ and dashed gray lines show smoothed means for $\alpha = 2$. The smoothed means were calculated using cubic smoothing splines.}
    \label{fig:CIratios}
\end{figure}

Figure~\ref{fig:CIratios} shows the widths of confidence intervals for $\widetilde{\beta}_\text{inf}$ versus $\hat{\beta}_\text{inf}$, $\widetilde{\beta}_\text{sus}$ versus $\hat{\beta}_\text{sus}$, $\widetilde{\beta}_\text{pair}$ versus $\hat{\beta}_\text{pair}$, and $\widetilde{\Lambda}_0(\tau)$ versus $\hat{\Lambda}_0(\tau)$. Knowledge of who-infects-whom improves the precision of $\beta_\text{inf}$ and $\beta_\text{pair}$ estimates but not $\beta_\text{sus}$ estimates; it slightly improves the precision of $\Lambda_0(\tau)$ estimates. The baseline hazard plays an important role in determining how much precision is gained, with a larger gain for $\alpha = 0.5$ than for $\alpha = 2$. The confidence intervals for $\widetilde{\beta}_\text{inf}$, $\widetilde{\beta}_\text{sus}$, $\widetilde{\beta}_\text{pair}$, and $\widetilde{\Lambda}_0(\tau)$ have slightly lower coverage probabilities than those for $\hat{\beta}_\text{inf}$, $\hat{\beta}_\text{sus}$, $\hat{\beta}_\text{pair}$, and $\hat{\Lambda}_0(\tau)$ (see Tables~\ref{tab:betaCI} and~\ref{tab:LambdaCI}), so these plots underestimate the true precision gained when who-infects-whom is observed.

Knowledge of who-infected-whom allows point estimates that are closer to the truth and interval estimates with better coverage probabilities. However, it is remarkable how much information can be recovered by the EM algorithm when who-infected-whom is not observed, making the iterative regression model of Section~\ref{sec:tilde} a promising tool for infectious disease epidemiology.

\section{Data Analysis}
\label{sec:data}
To show how the methods of Section~\ref{sec:methods} can be applied, we will look at the effect of antiviral prophylaxis and age on the transmission of pandemic influenza A(H1N1) in Los Angeles County in 2009. The Los Angeles County Department of Public Health (LACDPH) collected household surveillance data between April 22 and May 19, 2009 according to the following protocol~\citep{SugimotoLA}:
\begin{enumerate}
    \item Nasopharyngeal swabs and aspirates were taken from individuals who reported to the LACDPH or other health care providers with acute febrile respiratory illness (AFRI), defined as a fever $\geq 100^\circ \text{F}$ plus cough, core throat, or runny nose. These specimens were tested for influenza, and the age, gender, and symptom onset date of the AFRI patient were recorded.
    \item Patients whose specimens tested positive for pandemic influenza A(H1N1) or for influenza A of undetermined subtype were enrolled as index cases. Each of them was given a structured phone interview to collect the following information about his or her household contacts: age, gender, type of contact (household, intimate, in-home daycare, non-home daycare), and high risk status (pregnant, child on long-term aspirin therapy, immunosuppressed, or history of a chronic cardiac, pulmonary, renal, liver, or neurologic condition). The interviewer also recorded whether prophylactic antiviral medication was being taken by the household contacts. They were asked to report the symptom onset date of any AFRI episodes among their household contacts. 
    \item When necessary, a follow-up interview was given 14 days after the symptom onset date of the index case to assess whether any additional AFRI episodes had occurred in the household, including their illness onset date.
\end{enumerate}
There were 58 households with a total of 299 members. There were 99 infections, of whom 62 were index cases (4 of the 58 households had co-primary cases) and 27 were household contacts with an AFRI. For simplicity, we assume these were all influenza A(H1N1) cases and that all household members were susceptible to infection.

Our natural history assumptions are adapted from~\citet{YangH1N1} and identical to those in~\citet{Kenah5}. In the primary analysis, we assume an incubation period of 2 days, a latent period of 0 days, and an infectious period of 6 days. Under these assumptions, a person $j$ with symptom onset at time $t_j^\text{sym}$ was infected at time $t_j = t_j^\text{sym} - 2$ and will stop being infectious at time $t_j + 6 = t_j^\text{sym} + 4$. Under these assumptions, person $j$ can transmit infection on days $t_j + 1$ to $t_j + 6$. In a sensitivity analysis, we vary the latent period from $0$ to $1$ days, and the infectious period from $5$ to $7$ days. 

Here, we use the regression model of Section~\ref{sec:tilde} to estimate the influenza transmission hazard ratios for age in the infectious and the susceptible and the hazard ratio for antiviral prophylaxis in the susceptible. We then estimate transmission probabilities for different combinations of covariates in infectious/susceptible pairs. The variables in the regression models are: age$_\text{inf} = 0$ if the infectious person is $< 18$ years old and $1$ otherwise, age$_\text{sus} = 0$ if the susceptible is $< 18$ years old and $1$ otherwise, and proph$_\text{sus} = 0$ if the susceptible is not on antiviral prophylaxis and $1$ otherwise. Since antiviral prophylaxis was initiated after the initial case in each household, it was considered only as a susceptibility covariate. All statistical analysis was done in R 2.15 (\texttt{www.r-project.org}).

\subsection{Results}
There were $114$ people aged $< 18$ years and $185$ aged $\geq 18$ years, with no missing age data. There were $91$ people taking antiviral prophylaxis and $152$ not taking prophylaxis, with missing prophylaxis data for $56$ people. When who-infects-whom is not observed, a complete-case analysis requires the removal of all rows corresponding to infectious-susceptible pairs $ij$ where $i \in \mathcal{V}_j$ and any member of $\mathcal{V}_j$ is missing data. Otherwise, the remaining members of $\mathcal{V}_j$ get too much credit for the infection of $j$.

In the main analysis, there were $70$ people infected from outside the household (i.e., no possible infector in the household), $16$ with $1$ possible infector, $7$ with $2$ possible infectors, $4$ with $4$ possible infectors, and $2$ with $8$ possible infectors, giving us $1^{16} \times 2^7 \times 4^4 \times 8^2 = 2097152$ possible transmission trees. The pairwise data contains $443$ infectious-susceptible pairs with a total of $2455$ pair-days at risk of infection. Of these, $16 \times 1 + 7 \times 2 + 4 \times 4 + 2 \times 8 = 62$ rows represent possible infection events. All models used the Efron approximation for the partial likelihood with tied failure times.

The top panel of Table~\ref{tab:LAmodels} shows the results of seven models. All of the models including prophylaxis suggested that antiviral prophylaxis reduced the hazard of transmission by about 60\%, with low p-values. Multivariable and stratified models with interaction suggest a stronger effect of antiviral prophylaxis on transmission to and from adults than on transmission to and from children. However, the interaction term coefficients had high p-values and wide confidence intervals (not shown). In all models, adults appeared more infectious and less susceptible than children. However, the coefficients for the main effect of age also had high p-values and wide confidence intervals. The bottom panel of Table~\ref{tab:LAmodels} shows the results of a sensitivity analysis with the multivariable model without interaction. Varying the latent and infectious periods has relatively little effect on the results of the model.

\begin{figure}[p]
    \includegraphics[width = \textwidth]{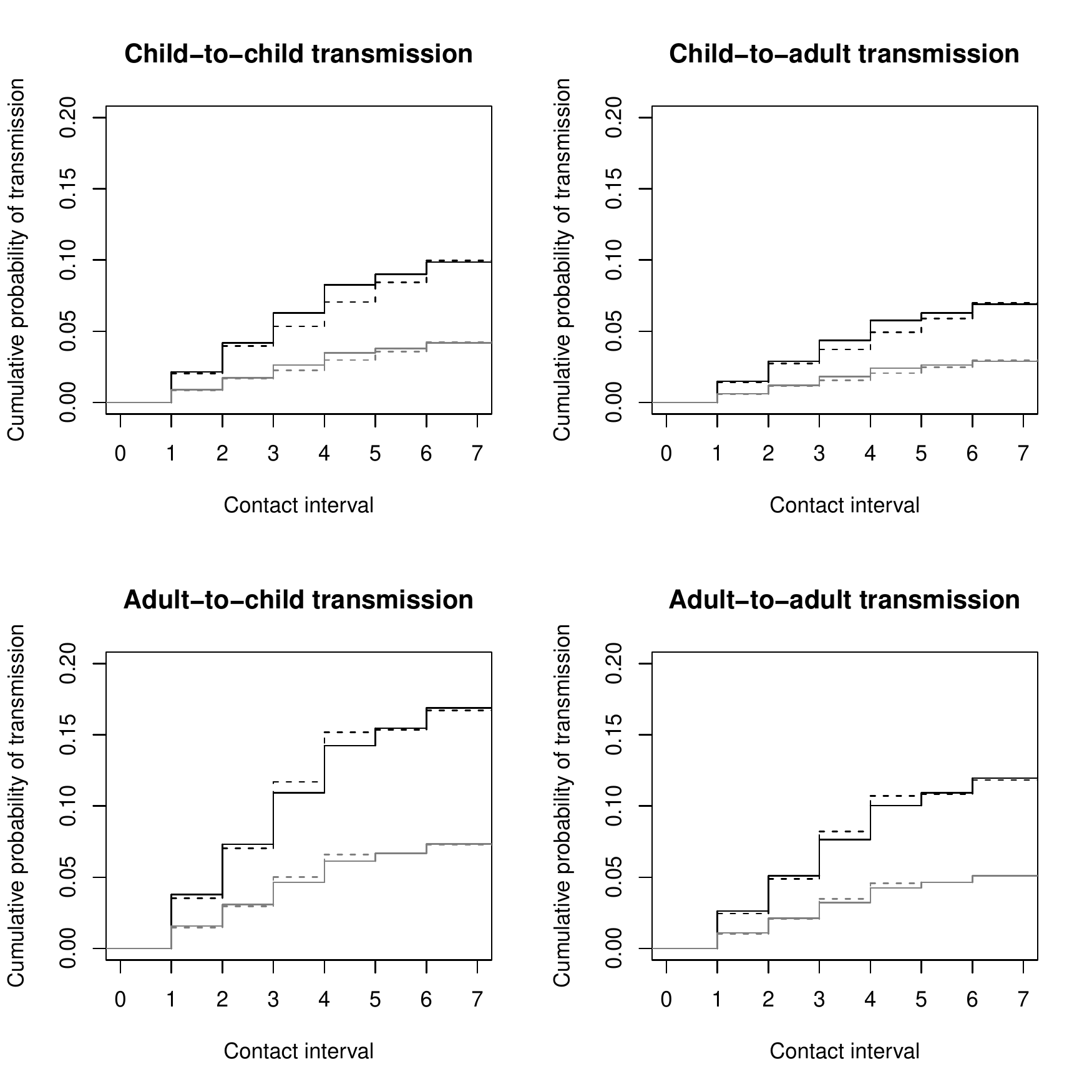}
    \caption{Household transmission of 2009 pandemic influenza A(H1N1) in Los Angeles County. Each panel shows separate curves for susceptible contacts with (gray lines) and without (black lines) antiviral prophylaxis. The solid lines are based on the multivariable model without interaction. The dotted lines are based on the model stratified by age$_\text{inf}$ without interaction.}
    \label{fig:LAh1n1}
\end{figure}

Figure~\ref{fig:LAh1n1} shows estimates of the cumulative transmission probability based on the multivariable and stratified models without interaction. The results of the two models are similar, but the stratified model generally showed slightly lower probabilities of transmission from children and higher probabilities of transmission from adults than the multivariable model. All four panels clearly show the estimtated effect of antiviral prophylaxis. Comparing the top and bottom rows shows that children are estimated to be less infectious than adults. Comparing the left and right columns shows that children are estimated to be more susceptible than adults. All curves show bigger jumps on the first four days after infection than on days $5$ and $6$, which is consistent with the results of~\citet{Kenah5}.

This data analysis has been intended primarily to illustrate the flexibility of the regression modeling framework for the analysis transmission data. There are several important limitations of the analysis itself. The data set is not large, so there is limited power to estimate the effects of age and antiviral prophylaxis. The age classification is crude, so it may not accurately capture the true effects of age. The prophylaxis variable was missing for many pairs and it was modeled as a binary variable, allowing no consideration of the timing of prophylaxis relative to exposure. Earlier analyses of household transmission of influenza A(H3N2) found greater child-to-child than adult-to-adult transmission~\citep{Addy}. In our analysis of influenza A(H1N1), we found that children are less infectious and more susceptible than adults. This could be a difference between the H3N2 and H1N1 subtypes of influenza A, or it could be a bias caused the failure to account for infection from outside the household. In either case, this analysis shows the need for two important extensions to the modeling framework: The ability to handle missing data more flexibly and the ability to model infection from outside the household.

\section{Discussion}
\label{sec:discussion}
Compared to the discrete-time chain binomial model of~\citet{RampeyLongini}, the regression model framework proposed here has several advantages. It can be fit using standard statistical software in a way that resembles a standard regression model. It offers all of the modeling tools available in a multivariable Cox regression framework, such as stratification and interaction. Standard software can be used to convert the results into curves representing the cumulative probability of transmission in pairs of individuals with specific characteristics. This ease of use will encourage the adoption of these methods in biomedical and public health research. 

There are two immediate extensions that will be required before the relative-risk regression models presented here can become truly useful tools in infectious disease epidemiology. First, we must be able to simultaneously model the process of infection from outside the household and transmission within the household. The discrete-time chain binomial model can include a per-time-unit probability of escaping infection from outside the household. In the iterative regression model, this could be achieved by fitting two models in each step of the EM algorithm: a pairwise contact interval model within the household and an individual-level absolute-time model for infection from outside the household. In iteration $k$ of the EM algorithm, an individual $j$ who got infected would have a probability $p_{0j}^{(k)}$ that he or she was infected from outside the household. The weights of the possible infectors within the household would add up to $1- p_{0j}^{(k)}$. At each step, the weights would be recalculated based on covariates, coefficient estimates, the baseline hazard of the contact interval distribution, and the baseline hazard of infection from outside the household. Second, we must be able to handle missing data flexibly but rigorously. Missing data on infection times, latent periods, and infectious periods is the rule, not the exception, in infectious disease epidemiology. For simple missing data (such as the missing data on antiviral prophylaxis in Section~\ref{sec:data}), the EM algorithm could be extended to calculate the expected log likelihood over the possible values of the missing data as well as who-infected-whom. A more general solution for missing data, especially missing infection and removal times, would be to use a profile sampler~\citep{LeeKosorokFine} for the model coefficients, treating the baseline hazards as a nuisance parameter.

Other extenions of the theory and methods presented here would make the regression framework presented here more broadly applicable. The SEIR framework is best suited to acute, immunizing diseases that spread directly from person to person. Many foodborne and waterborne diseases, pneumococcal and meningococcal diseases, and other infectious diseases of major public health importance do not fit easily into this framework. The first limitation could be addressed by allowing individuals to experience multiple events (first infection, the second infection, etc.) and allowing individuals to experience different types of events (new carriage, new infection, relapse, etc.). In this paper, we assumed that contact intervals are independent of infectious periods. In some cases, there may be a covariate process $X(\tau)$ such that $I_i(\tau)$ and $\mathcal{N}_{ij}(\tau)$ are independent given $X(\tau^-)$. If not, infectious contact and the infectious period could be modeled as a multivariate survival process. The flexibility of the theory of counting processes and martingales will be valuable in extending the model to more complex diseases.

Finally, there are technical issues that deserve more study. The smoothing step is crucial to the iterative regression model. Here, we used cubic smoothing splines because they were convenient and worked well. However, these do not guarantee that the smoothed hazard function is monotonically increasing and do not have a convenient interpretation in terms of the likelihood. A penalized likelihood estimator that guarantees monotonicity~\citep{AndersonSenthilselvan} would be more consistent with the theoretical justification of the EM algorithm. A more careful study of the asymptotics of this model would also be useful, especially as the model is extended to more complex applications. Since most infectious disease data is discrete (by day, by week, etc.), a detailed comparison of regression models with correction for ties versus the discrete-time chain-binomial model is important.

Despite these limitations, semiparametric relative-risk regression is a powerful new framework for the analysis of infectious disease data. Its flexibility will allow statistical methods in infectious disease epidemiology to develop in concert with advances in molecular biology. Since all calculations in the EM algorithm are sums over possible combinations of who-infected whom, phylogenetic data can be incorporated directly by restricting the sums to transmission trees compatible with the phylogenetic tree. By placing the analysis of infectious disease data on the theoretical foundation of survival analysis, this approach may help clarify causal inference in infectious disease epidemiology, allowing better design of observational studies and intervention trials. Statistical methods that help improve the response to emerging or re-emerging infections could protect human health and commerce from unknown but possibly tremendous dangers.

\bibliographystyle{imsart-nameyear}
\bibliography{spCIregression}

\appendix
\section{Consistency and asymptotic normality}
\label{app:asymptotics}
The conditions for the consistency and asymptotic normality of $\hat{\beta}$ and $\hat{\Lambda}_0(\tau)$ in the Cox model were given in~\citet{AndersenGill1982}, which used martingales to simplify and generalize the asymptotic results of~\citet{Cox1975} and~\citet{Tsiatis1981}. Conditions for the more general relative risk model were given in~\citet{PrenticeSelf}. Here, we outline the most important of these conditions and point out their implications for the use of relative risk regression models in infectious disease epidemiology. 

\subsection{Regularity conditions}
Assume all observations take place at infectiousness ages in $[0, \mathcal{T}]$ for some finite $\mathcal{T}$. Let $m = Y(0^+) = \lim_{\tau \downarrow 0} Y(\tau)$ be the number of pairs $ij$ that were at risk of infectious contact from $i$ to $j$ while under observation. Let $n_m$ denote the number of individuals that constitute the $m$ pairs. Define the following functions~\citep{PrenticeSelf}:
\begin{align*}
    S^{(0)}_m(\beta, \tau) &= \frac{1}{m} Y(\beta, \tau) = \sum_{j=1}^{n_m} \sum_{i\neq j} r\big(\beta^\trans X_{ij}(\tau)\big) Y_{ij}(\tau), \\
    S^{(1)}_m(\beta, \tau) &= \deriv{\beta} S^{(0)}_m(\tau) = \frac{1}{m} \sum_{j=1}^{n_m} X_{ij}(\tau) r'\big(\beta^\trans X_{ij}(\tau)\big) Y_{ij}(\tau), \text{ and} \\
    S^{(2)}_m(\beta, \tau) &= \frac{1}{m} \sum_{j=1}^{n_m} \sum_{i\neq j} X_{ij}(\tau)^{\otimes 2} \Big((\ln r)'\big(\beta^\trans X_{ij}(\tau)\big)\Big)^2 r\big(\beta^\trans X_{ij}(\tau)\big) Y_{ij}(\tau). \\
\end{align*}
Note that $S^{(0)}_m$ is real-valued, $S^{(1)}_m$ is $b\times 1$ vector-valued, and $S^{(2)}_m$ is $b \times b$ matrix-valued. Now let
\begin{align}
    E_m(\beta, \tau) &= \frac{S^{(1)}_m(\beta, \tau)}{S^{(0)}_m(\beta, \tau)} \text{ and } \\
    V_m(\beta, \tau) &= \frac{S^{(2)}_m(\beta, \tau)}{S^{(0)}_m(\beta, \tau)} - E_m(\beta, \tau)^{\otimes 2}
\end{align}
be the values of $E(\beta, \tau)$ and $V(\beta, \tau)$, respectively, based on observations of $m$ pairs at risk of transmission.

For consistency and asymptotic normality of $\sqrt{m}(\hat{\beta} - \beta_0)$, we have the following sufficient conditions~\citep{AndersenGill1982, PrenticeSelf}:
\begin{enumerate}[A.]
    \item \label{finite} (Finite interval) $\Lambda_0(\mathcal{T}) < \infty$.
    \item \label{rpos} (Regression function positivity) There exists a neighborhood $\mathcal{B}_0$ of $\beta_0$ such that $r\big(\beta^\trans X_{ij}(\tau)\big)$ is locally bounded away from zero for all $ij$ and all $\beta \in \mathcal{B}_0$. 
    \item (Asymptotic stability) There exists a neighborhood $\mathcal{B} \subseteq \mathcal{B}_0$ of $\beta_0$ and functions $s^{(0)}, s^{(1)}, s^{(2)}$ defined on $\mathcal{B} \times [0, \mathcal{T}]$ such that 
        \begin{equation}
            \sup_{\beta \in \mathcal{B}, \tau \in [0, \mathcal{T}]} \|S^{(k)}_m(\beta, \tau) - s^{(k)}(\beta, \tau)\| \prob 0 \text{ as } m \rightarrow \infty
        \end{equation}
        for $k = 0, 1, 2$. Here, $\|x\|$ is $|x|$ for real $x$, $\max\big(|x_1|,\ldots, |x_b|\big)$ for vector $x$, and $\max\big(|x_{11}|,\ldots, |x_{bb}|\big)$ for matrix $x$. Asymptotic properties of the Cox model depend only on convergence of these three functions. For more general relative risk functions, convergence of four additional functions is also required~\citep{PrenticeSelf}.
    \item (Asymptotic regularity) The functions $s^{(0)}(\beta, \tau), \ldots, s^{(2)}(\beta, \tau)$ are bounded on $\mathcal{B} \times [0, \mathcal{T}]$ and continuous in $\beta$ uniformly in $\tau$. In addition, $s^{(0)}$ is bounded away from zero and has first and second derivatives with respect to $\beta$ on $\mathcal{B} \times [0, \mathcal{T}]$. Finally, let $e(\beta, \tau) = \frac{s^{(1)}(\beta, \tau)}{s^{(0)}(\beta, \tau)}$ and $v(\beta, \tau) = \frac{s^{(2)}(\beta, \tau)}{s^{(0)}(\beta, \tau)} - e(\beta, \tau)^{\otimes 2}$. Then
        \begin{equation}
            \Sigma = \int_0^\mathcal{T} v(\beta_0, u) s^{(0)}(\beta_0, u) \lambda_0(u) \dif u
        \end{equation}
    is positive definite.
    \item \label{Istability} (Asymptotic stability of the observed information matrix) 
        \begin{multline}
            \sup_{\beta \in \mathcal{B}} \int_0^\mathcal{T} \frac{1}{m^2} \sum_{j=1}^{n_m} \sum_{i \neq j} \|X_{ij}(u)\|^4 \Big((\ln r)''\big(\beta^\trans X_{ij}(u)\big)\Big)^2 r\big(\beta_0^\trans X_{ij}(u)\big) \lambda_0(u) \dif u \\
            \prob 0.
        \end{multline}
    \item \label{Lindeberg} (Lindeberg condition)
        \begin{equation}
            \frac{1}{\sqrt{m}} \sup_{\tau, ij} \Big\|X_{ij}(\tau) (\ln r)'\big(\beta_0^\trans X_{ij}(\tau)\big)\Big\| \prob 0,
        \end{equation}
        where the supremum is over all $\tau \in [0, \mathcal{T}]$ and all $ij$ such that $Y_{ij}(\tau) = 1$.
\end{enumerate}
Condition~\ref{Lindeberg} is automatically fulfilled if the covariates $X_{ij}$ are bounded. In the Cox model, conditions~\ref{rpos} and~\ref{Istability} are automatically fulfilled because $\exp(x) > 0$ and $(\ln r)''(x) = 0$ for all real $x$. With only slight modification, these conditions also guarantee consistency and asymptotic normality in a stratified relative-risk regression model~\citep{AndersenBorgan1985}.

For the methods in this paper, the most important constraint is that $s^{(0)}(\beta, \tau)$ is bounded away from zero. This has two implications for infectious disease data that have no counterpart in standard survival data. First, the infectious period must be $\geq \mathcal{T}$ with positive probability. Second, the number of infectors to which the susceptible $j$ in a randomly chosen pair $ij$ at risk of transmission is exposed must have a finite mean as $m \rightarrow \infty$. Let 
\begin{equation}
    Y_{\cdot j}(\tau) = \sum_{i \neq j} Y_{ij}(\tau) \Rightarrow \sum_{j = 1}^{n_m} Y_{\cdot j}(0^+) = m.
\end{equation}
Now let $D_{ij} = Y_{\cdot j}(0^+) Y_{ij}(0^+)$ be the number of infectors to which $j$ is exposed if $ij$ was at risk of transmission and $D_{ij} = 0$ otherwise. If we randomly choose a pair $ij$ at risk of transmission and look at the number of infectors to which $j$ is exposed, its expected value is 
\begin{equation}
    \frac{1}{m} \sum_{j = 1}^{n_m} \sum_{i \neq j} D_{ij} = \frac{1}{m} \sum_{j=1}^{n_m} Y_{\cdot j}(0^+)^2.
\end{equation}
For $s^{(0)}(\beta, \tau)$ to be bounded away from zero, we must have
\begin{equation}
    \limsup_{m \rightarrow \infty} \frac{1}{m} \sum_{j = 1}^{n_m} Y_{\cdot j}(0^+)^2 < \infty.
\end{equation}
If not, the hazard of infection in $j$ from a randomly chosen $ij$ at risk of transmission becomes infinite as $m \rightarrow \infty$. Each susceptible will be infected at a contact interval approaching zero, so the contact interval distribution cannot be estimated. In practice, this means that large-sample distributions are useful when the number of pairs $m$ and the number of susceptibles are both large and the largest value of $Y_{\cdot j}(0^+) \ll m$.

There is no similar constraint on the number of susceptibles exposed to each infectious person. In theory, we could have $m$ susceptibles exposed to a single infectious person without violating the regularity conditions (as long as his or her infectious period was $\geq \mathcal{T}$). This is because the contact intervals in all pairs $ij$ for a fixed $i$ are assumed to be independent of each other and independent of the infectious period of $i$ conditional on the covariate processes $X_{ij}(\tau)$.

\subsection{Asymptotic properties of $U(\beta_0, \tau)$, $\hat{\beta}$, and $\hat{\Lambda}_0(\tau)$}
Let $U_m(\beta_0, \tau)$ denote the score process based on observations of $m$ pairs $ij$ at risk of transmission when who-infects-whom is observed, let $\hat{\beta}_m$ denote the corresponding maximum partial likelihood estimate, and let $\hat{\Lambda}_{0, m}(\tau)$ denote the corresponding Breslow estimate of the baseline hazard. Under the conditions of the last section, we have the following results as $m \rightarrow \infty$~\citep{AndersenGill1982, PrenticeSelf}:
\begin{enumerate}
    \item Asymptotic normality of the score: $\frac{1}{\sqrt{m}} U(\beta_0, \mathcal{T}) \dist N\big(0, \Sigma\big)$.
    \item Consistency of $I(\beta_0)$ and $\mathcal{I}(\beta_0)$: $\frac{1}{m}I(\beta_0) \prob \Sigma$ and $\frac{1}{m}\mathcal{I}(\beta_0) \prob \Sigma$.
    \item Consistency of $\hat{\beta}$: $\hat{\beta}_m \prob \beta_0$.
    \item Asymptotic normality of $\hat{\beta}$: $\sqrt{m}(\hat{\beta} - \beta_0) \dist N\big(0, \Sigma^{-1}\big)$.
    \item Consistency of $I(\hat{\beta})$ and $\mathcal{I}(\hat{\beta})$: $\frac{1}{m} I(\hat{\beta}) \prob \Sigma$ and $\frac{1}{m} \mathcal{I}(\hat{\beta}) \prob \Sigma$.
    \item Convergence of $\sqrt{m}\big(\hat{\Lambda}_0(\tau) - \Lambda_0(\tau)\big)$ to a mean-zero Gaussian process with independent increments.
    \item Asymptotic independence of $\Big(\deriv{\beta} \hat{\Lambda}(\beta^*, \tau)\Big)^\trans \sqrt{m} (\hat{\beta} - \beta_0)$ and $m \int_0^\tau \frac{1}{Y(\beta_0, u)^2} \dif N(u)$.
    \item Continuity of $\deriv{\beta} \hat{\Lambda}(\beta^*, \tau)$: $\deriv{\beta} \hat{\Lambda}(\beta_m, \tau) \prob \deriv{\beta} \hat{\Lambda}(\beta_0, \tau)$ if $\beta_m \prob \beta$.
\end{enumerate}

\section{Variance of baseline hazard estimates}
\citet{AndersenGill1982} showed that $\sqrt{m}\big(\hat{\Lambda}_0(\tau) - \Lambda_0(\tau)\big)$ converges to a mean-zero Gaussian martingale in the Cox model for standard survival data, and this result was extended to more general relative risk functions by~\citet{PrenticeSelf}. Under the conditions given in Appendix~\ref{app:asymptotics}, these derivations extend directly to infectious disease data.

\subsection{Who-infects-whom is observed}
\label{app:hatVariance}
Expanding $\hat{\Lambda}_0(\tau) - \Lambda_0(\tau)$ gives us
\begin{align}
    \sqrt{m} \Big(\hat{\Lambda}_0(\tau) - \Lambda_0(\tau)\Big) 
    &= \sqrt{m} \Big(\hat{\Lambda}(\hat{\beta}, \tau) - \hat{\Lambda}(\beta_0, \tau)\Big) \nonumber \\
    &\qquad + \sqrt{m} \Big(\hat{\Lambda}(\beta_0, \tau) - \Lambda_0^*(\tau)\Big) \nonumber \\
    &\qquad + \sqrt{m} \big(\Lambda_0^*(\tau) - \Lambda_0(\tau)\big), 
    \label{eq:varLambda}
\end{align}
where $\Lambda_0^*(\tau) = \int_0^\tau \indicator{Y(u) > 0} \lambda_0(u) \dif u$. By a first-order Taylor expansion, the first term in~\eqref{eq:varLambda} is
\begin{equation}
    \bigg(\deriv{\beta} \hat{\Lambda}(\beta^*, \tau)\bigg)^\trans \sqrt{m} \big(\hat{\beta} - \beta_0 \big)
\end{equation}
for some $\beta^*$ on the line segment between $\beta_0$ and $\hat{\beta}$. Using the Doob-Meyer decomposition, the second term in~\eqref{eq:varLambda} can be written
\begin{equation}
    \sqrt{m} \int_0^\tau \frac{\indicator{Y(u) > 0}}{Y(\beta_0, u)} \dif M(u),
\end{equation}
which is a martingale with the optional variation process
\begin{equation}
    m \int_0^\tau \frac{1}{Y(\beta_0, u)^2} \dif N(u).
\end{equation}
The third term in~\eqref{eq:varLambda} is zero For all $\tau$ such that $Y(\tau) > 0$. Under the regularity conditions of Appendix~\ref{app:asymptotics}, the first and second terms are asymptotically independent, so the asymptotic variance of~\eqref{eq:varLambda} is
\begin{equation}
     \bigg(\deriv{\beta} \hat{\Lambda}(\hat{\beta}, \tau)\bigg)^\trans \bigg(\frac{1}{m} I(\hat{\beta})\bigg)^{-1} \bigg(\deriv{\beta} \hat{\Lambda}(\hat{\beta}, \tau)\bigg) + \int_0^\tau \frac{m}{Y(\hat{\beta}, u)^2} \dif N(u)
\end{equation}
for all $\tau$ such that $Y(\tau) > 0$. 

\subsection{Who-infects-whom is not observed}
\label{app:tildeVariance}
By an expansion similar to that in equation~\eqref{eq:varLambda},
\begin{align}
    \sqrt{m} \Big(\widetilde{\Lambda}_0(\tau) - \Lambda_0(\tau)\Big) 
    &= \sqrt{m} \Big(\widetilde{\Lambda}_{\widetilde{\beta}, \widetilde{\lambda}}(\widetilde{\beta}, \tau) - \widetilde{\Lambda}_{\widetilde{\beta}, \widetilde{\lambda}}(\beta_0, \tau)\Big) \nonumber \\
    &\qquad + \sqrt{m} \Big(\widetilde{\Lambda}_{\widetilde{\beta}, \widetilde{\lambda}}(\beta_0, \tau) - \widetilde{\Lambda}_{\beta_0, \lambda_0}(\beta_0, \tau)\Big) \nonumber \\
    &\qquad + \sqrt{m} \Big(\widetilde{\Lambda}_{\beta_0, \lambda_0}(\beta_0, \tau) - \Lambda_0^*(\tau)\Big) \nonumber \\
    &\qquad + \sqrt{m} \big(\Lambda_0^*(\tau) - \Lambda_0(\tau)\big). 
    \label{eq:varTildeLambda}
\end{align}
The fourth term in~\eqref{eq:varTildeLambda} is zero for all $\tau$ at which $Y(\tau) > 0$. 

By a first-order Taylor expansion, the first term in~\eqref{eq:varTildeLambda} equals 
\begin{equation}
    \bigg(\deriv{\beta} \widetilde{\Lambda}_{\widetilde{\beta}, \widetilde{\lambda}}(\beta^*, \tau)\bigg)^\trans \sqrt{m} \big(\widetilde{\beta} - \beta_0\big)
\end{equation}
for some $\beta^*$ on the line segment between $\beta_0$ and $\widetilde{\beta}$, where
\begin{equation}
    \deriv{\beta} \widetilde{\Lambda}_{\widetilde{\beta}, \widetilde{\lambda}}(\beta, \tau) = -\int_0^\tau \frac{\deriv{\beta} Y(\beta, u)}{Y(\beta, u)^2} \dif \widetilde{N}(u | \widetilde{\beta}, \widetilde{\lambda}).
\end{equation}
Its contribution to the variance is 
\begin{equation}
    \bigg(\deriv{\beta} \widetilde{\Lambda}_{\widetilde{\beta}, \widetilde{\lambda}}(\beta_0, \tau)\bigg)^\trans \bigg(\frac{1}{m} \widetilde{I}(\beta_0)\bigg)^{-1} \bigg(\deriv{\beta} \widetilde{\Lambda}_{\widetilde{\beta}, \widetilde{\lambda}}(\beta_0, \tau)\bigg).
\end{equation}

The second term in~\eqref{eq:varTildeLambda} can be rewritten
\begin{equation}
    \sqrt{m} \int_0^\tau \frac{1}{Y(\beta_0, u)} \Big(\dif[] \widetilde{N}(u | \widetilde{\beta}, \widetilde{\lambda}) - \dif[] \widetilde{N}(u | \beta_0, \lambda_0)\Big) 
\end{equation}
For each $j$, we have $\int_0^\infty \dif \widetilde{N}(u|\beta, \lambda) = 1$ if $j$ was infected and $0$ otherwise. Thus, the term in parentheses is the sum a subset of the random variables $\delta_{ij} = p_{ij}(\widetilde{\beta}, \widetilde{\lambda}) - p_{ij}(\beta_0, \lambda_0)$, which have sum zero for each $j$. Since the $\delta_{ij}$ are asymptotically independent for different $j$ and $Y(\beta_0, u) = O_P(m)$, the integral behaves asymptotically like a mean of independent random variables with mean zero and variance $O(\widetilde{\beta} - \beta_0)$. Therefore, the second term of~\eqref{eq:varTildeLambda} is $O_P(\widetilde{\beta} - \beta_0)$ and converges in probability to zero as $m \rightarrow \infty$. 

The third term in~\eqref{eq:varTildeLambda} can be evaluated using the conditional variance formula. The expression inside the parentheses has the variance
\begin{multline}
    \subE{\beta_0, \lambda_0}{\hat{\sigma}^2_\mathbf{v}(\beta_0, \tau)} + \subVar[\big]{\beta_0, \lambda_0}{\hat{\Lambda}_\mathbf{v}(\beta_0, \tau)} = \\
    \int_0^\tau \frac{1}{Y(\beta_0, u)^2} \dif \widetilde{N}(u | \beta_0, \lambda_0) + \subE[\big]{\beta_0, \lambda_0}{\hat{\Lambda}_\mathbf{v}(\beta_0, \tau)^2} - \widetilde{\Lambda}_{\beta_0, \lambda_0}(\beta_0, \tau)^2,
\end{multline}
where 
\begin{equation}
    \hat{\sigma}^2_\mathbf{v}(\beta, \tau) = \int_0^\tau \frac{1}{Y(\beta, u)^2} \dif N(u|\mathbf{v}).
\end{equation}
Since each infected person has only one infector and infectors can be chosen independently given the observed data,
\begin{align}
    \subE[\big]{\beta_0, \lambda_0}{\hat{\Lambda}_\mathbf{v}(\beta_0, \tau)^2} 
    &= \widetilde{\Lambda}_{\beta_0, \lambda_0}(\widetilde{\beta}, \tau)^2 - \sum_{j=1}^n \bigg(\int_0^\tau \frac{1}{Y(\beta_0, u)} \dif \widetilde{N}_{\cdot j}(u | \beta_0, \lambda_0)\bigg)^2 \nonumber \\
    &\qquad + \int_0^\tau \frac{1}{Y(\beta_0, u)^2} \dif \widetilde{N}(u | \beta_0, \lambda_0),
\end{align}
where $\widetilde{N}_{\cdot j}(u | \beta, \lambda) = \sum_{i \neq j} \widetilde{N}_{ij}(u | \beta, \lambda)$. Therefore, the total variance contribution of the third term in~\eqref{eq:varTildeLambda} reduces to
\begin{equation}
    2 \int_0^\tau \frac{m}{Y(\beta_0, u)^2} \dif \widetilde{N}(u | \beta_0, \lambda_0) - \sum_{j=1}^n \bigg(\int_0^\tau \frac{\sqrt{m}}{Y(\beta_0, u)} \dif \widetilde{N}_{\cdot j}(u | \beta_0, \lambda_0)\bigg)^2.
    \label{eq:var3}
\end{equation}

Since only the first and third terms of~\eqref{eq:varTildeLambda} are asymptotically nonzero, all that remains is to look at their covariance. Let $N_{ij}(\tau|\mathbf{v})$ denote the value of $N_{ij}(\tau)$ that we would have calculated had we observed the transmission network $\mathbf{v}$. Then the corresponding value of the score $U(\beta, \tau)$ is  
\begin{equation}
    U_\mathbf{v}(\beta, \tau) = \sum_{j=1}^n \sum_{i\neq j} \int_0^\tau \deriv{\beta} \ln \frac{r\big(\beta^\trans X_{ij}(u)\big)}{Y(\beta, u)} \dif N(u|\mathbf{v})
\end{equation}
and the corresponding covariance of $U(\beta, \tau)$ and $\hat{\Lambda}(\beta, \tau)$ is
\begin{equation}
    \kappa_\mathbf{v}(\beta, \tau) = \sum_{j=1}^n \sum_{i\neq j} \int_0^\tau \frac{1}{Y(\beta, u)} \bigg(\deriv{\beta} \ln \frac{r\big(\beta^\trans X_{ij}(u)\big)}{Y(\beta, u)}\bigg) \dif N_{ij}(u|\mathbf{v}).
\end{equation}
By the conditional covariance formula,
\begin{align}
    \Cov[\Big]{\widetilde{U}_{\beta_0, \lambda_0}(\beta_0, \lambda_0)}{\widetilde{\Lambda}_{\beta_0, \lambda_0}(\beta_0, \tau)} 
    &= \subCov[\big]{\beta_0, \lambda_0}{U_\mathbf{v}(\beta_0, \tau)}{\hat{\Lambda}_\mathbf{v}(\beta_0, \tau)} \nonumber \\
    &\qquad + \subE{\beta_0, \lambda_0}{\kappa_\mathbf{v}(\beta_0, \tau)}
\end{align}
By an argument similar to that leading to~\eqref{eq:var3}, this reduces to
\begin{align}
    &2 \sum_{j=1}^n \sum_{i\neq j} \int_0^\tau \frac{1}{Y(\beta_0, u)} \bigg(\deriv{\beta} \ln \frac{r\big(\beta_0^\trans X_{ij}(u)\big)}{Y(\beta_0, u)}\bigg) \dif \widetilde{N}_{ij}(u|\beta_0, \lambda_0) \nonumber \\
    &\qquad - \sum_{j=1}^n \bigg(\int_0^\tau \frac{1}{Y(\beta_0, u)} \dif \widetilde{N}_{\cdot j}(u|\beta_0, \lambda_0)\bigg) U_{\cdot j}(\beta_0, \tau).
    \label{eq:cov13}
\end{align}
In the limit of large $m$, both terms in~\eqref{eq:cov13} act like means of random variables with mean zero and finite variance, so they converge in probability to zero. Since $\widetilde{\beta}$ is a function of the expected score, this implies that the first and third terms of equation~\eqref{eq:varTildeLambda} are asymptotically independent. 

Combining all of these results, the asymptotic variance of~\eqref{eq:varTildeLambda} is
\begin{multline}
    \bigg(\deriv{\beta} \widetilde{\Lambda}_{\widetilde{\beta}, \widetilde{\lambda}}(\beta_0, \tau)\bigg)^\trans \bigg(\frac{1}{m} \widetilde{I}(\beta_0)\bigg)^{-1} \bigg(\deriv{\beta} \widetilde{\Lambda}_{\widetilde{\beta}, \widetilde{\lambda}}(\beta_0, \tau)\bigg) \\
    + 2 \int_0^\tau \frac{m}{Y(\beta_0, u)^2} \dif \widetilde{N}(u | \beta_0, \lambda_0)
    - \sum_{j=1}^n \bigg(\int_0^\tau \frac{\sqrt{m}}{Y(\beta_0, u)} \dif \widetilde{N}_{\cdot j}(u | \beta_0, \lambda_0)\bigg)^2.
\end{multline}

\clearpage

\begin{table}[p]
    \centering
    \begin{tabular}{c|cccc}
    & \multicolumn{4}{c}{Parameter: $\beta_\text{inf}$} \\
    & \multicolumn{2}{c}{$\beta_\text{sus} = \beta_\text{pair} = 0$} & \multicolumn{2}{c}{$\beta_\text{sus} = \beta_\text{pair} = 1$} \\
    Baseline hazard & $\hat{\beta}_\text{inf}$ & $\widetilde{\beta}_\text{inf}$ & $\hat{\beta}_\text{inf}$ & $\widetilde{\beta}_\text{inf}$ \\
    \hline
    $\alpha = .5$ & .958 & .956 & .945 & .940 \\
    $\alpha = 2$ & .939 & .932 & .938 & .920 \\
    \multicolumn{5}{c}{} \\
    & \multicolumn{4}{c}{Parameter: $\beta_\text{sus}$} \\
    & \multicolumn{2}{c}{$\beta_\text{inf} = \beta_\text{pair} = 0$} & \multicolumn{2}{c}{$\beta_\text{inf} = \beta_\text{pair} = 1$} \\
    Baseline hazard & $\hat{\beta}_\text{sus}$ & $\widetilde{\beta}_\text{sus}$ & $\hat{\beta}_\text{sus}$ & $\widetilde{\beta}_\text{sus}$ \\
    \hline
    $\alpha = .5$ & .945 & .943 & .946 & .952 \\
    $\alpha = 2$ & .918 & .921 & .932 & .932 \\
    \multicolumn{5}{c}{} \\
    & \multicolumn{4}{c}{Parameter: $\beta_\text{pair}$} \\
    & \multicolumn{2}{c}{$\beta_\text{inf} = \beta_\text{sus} = 0$} & \multicolumn{2}{c}{$\beta_\text{inf} = \beta_\text{sus} = 1$} \\
    Baseline hazard & $\hat{\beta}_\text{pair}$ & $\widetilde{\beta}_\text{pair}$ & $\hat{\beta}_\text{pair}$ & $\widetilde{\beta}_\text{pair}$ \\
    \hline
    $\alpha = .5$ & .949 & .928 & .952 & .940 \\
    $\alpha = 2$ & .950 & .941 & .951 & .934
    \end{tabular}
    \caption{95\% confidence interval coverage probabilities in simulations. Each probability is based on the results of $1000$ simulations.}
    \label{tab:betaCI}
\end{table}

\begin{table}[p]
    \centering
    \begin{tabular}{c|cccc}
        Baseline hazard & \multicolumn{2}{c}{$\alpha = .5$} & \multicolumn{2}{c}{$\alpha = 2$} \\
        Quantile & $\hat{\Lambda}_0(\tau)$ & $\widetilde{\Lambda}_0(\tau)$ & $\hat{\Lambda}_0(\tau)$ & $\widetilde{\Lambda}_0(\tau)$ \\
        \hline
        10\% & .944 & .925 & .956 & .846 \\
        25\% & .947 & .923 & .942 & .809 \\
        50\% & .946 & .924 & .930 & .792 \\
        75\% & .945 & .917 & .908 & .793 \\
        90\% & .942 & .922 & .887 & .797
    \end{tabular}
    \caption{95\% confidence interval coverage probabilities in simulations. Each probability is based on the results of $6000$ simulations.}
    \label{tab:LambdaCI}
\end{table}

\begin{table}[p]
    \centering
    \makebox[\textwidth]{
    \begin{tabular}{lccccc}
        & \multicolumn{3}{c}{\textbf{Covariates}} & \\
        \textbf{Model} & age$_\text{inf}$ & age$_\text{sus}$ & prophy$_\text{sus}$ & \multicolumn{2}{c}{Interaction terms} \\ \hline \\
        Univariable & 1.53 (0.66, 3.54) & 0.41 (0.20, 0.85) & 0.43 (0.18, 1.02) & & \\
        & $p = .321$ & $p = .016$ & $p = .057$ & & \\ \\
        Multivariable & 1.78 (0.69, 4.62) & 0.69 (0.29, 1.64) & 0.41 (0.17, 0.98) & & \\ 
        & $p = .234$ & $p = .399$ & $p = .046$ & \\ \\
        & & & & age$_\text{inf}$:age$_\text{sus}$ & 0.66 ($p = .705$) \\
        Multivariable & 1.59 (0.32, 7.84) & 0.63 (0.14, 2.73) & 0.04 (0.00, 9.62) & age$_\text{inf}$:proph$_\text{sus}$ & 9.28 ($p = .450$)  \\ 
        + interaction & $p = .570$ & $p = .532$ & $p = .253$ & age$_\text{sus}$:proph$_\text{sus}$ & 2.72 ($p = .361$) \\
        & & & & Likelihood ratio & $p = .101$ \\ \\ 
        Stratified & strata &  0.69 (0.29, 1.64) & 0.41 (0.17, 0.99) & & \\ 
        & & $p = .401$ & $p = .047$ & & \\ \\
        Stratified & strata & 0.52 (0.29, 1.64) & 0.23 (0.05, 1.16) & age$_\text{inf}$:proph$_\text{sus}$ & 2.37 ($p = .379$) \\
        + interaction & & $p = .219$ & $p = .075$ & Likelihood ratio & $p = .353$ \\ \\
        \hline \\
        \multicolumn{6}{c}{\textbf{Sensitivity analysis (multivariable model without interaction)}} \\
        Latent period & & & & & \\
        $1$ day & 1.44 (0.64, 3.26) & 0.83 (0.36, 1.93) & 0.35 (0.15, 0.80) & & \\
        & $p = .378$ & $p = .670$ & $p = .013$ \\
        Infectious period & & & & & \\
        $5$ days & 1.59 (0.60, 4.20) & 0.64 (0.27, 1.55) & 0.45 (0.18, 1.07) & & \\
        & $p = .348$ & $p = .322$ & $p = .073$ & & \\
        $7$ days & 1.45 (0.62, 3.40) & 0.89 (0.38, 2.04) & 0.34 (0.17, 0.87) & & \\
        & $p = .378$ & $p = .670$ & $p = .013$ & &
    \end{tabular}
    }
    \caption{Hazard ratios and p-values for different models of the 2009 pandemic influenza A(H1N1) household surveillance data from Los Angeles County. The multivariable and stratified models without interaction were used as the final models.}
    \label{tab:LAmodels}
\end{table}

\end{document}